\documentclass[twoside]{article}
\usepackage{qic,epsfig}
\usepackage{eufrak}

\newcommand{\ket}[1]{\left\vert{#1}\right\rangle}

\begin{document}
\setlength{\textheight}{8.0truein}    

\runninghead{A scheme to protect against multiple quantum erasures}
            {Gilson O. dos Santos and Francisco M. de Assis}

\normalsize\textlineskip
\thispagestyle{empty}
\setcounter{page}{1}


\vspace*{0.88truein}

\alphfootnote

\fpage{1}

\centerline{\bf
A SCHEME TO PROTECT AGAINST MULTIPLE QUANTUM ERASURES}
\vspace*{0.035truein}
\vspace*{0.37truein}
\centerline{\footnotesize
GILSON O. DOS SANTOS
}
\vspace*{0.015truein}
\centerline{\footnotesize\it Federal Institute of Education, Science and Technology of Alagoas}
\baselineskip=10pt
\centerline{\footnotesize\it Maceio, Alagoas, 57020-600, Brazil, E-mail: gilson.santos@ee.ufcg.edu.br}
\vspace*{10pt}
\centerline{\footnotesize 
FRANCISCO M. DE ASSIS}
\vspace*{0.015truein}
\centerline{\footnotesize\it Department of Electrical Engineering, Federal University of Campina Grande}
\baselineskip=10pt
\centerline{\footnotesize\it Campina Grande, Para\'{i}ba, 58429-140, Brazil, E-mail: fmarcos@dee.ufcg.edu.br}
\vspace*{0.225truein}

\vspace*{0.21truein}

\abstracts{
We present a scheme able to protect $k\geq 3$ qubits of information against the occurrence of multiple erasures, based on the code proposed by Yang et al. (2004 JETP Letters {\bf 79} 236). In this scheme redundant blocks are used and we restrict to the case that each erasure must occur in distinct blocks. We explicitly characterize the encoding operation and the restoring operation required to implement this scheme. The operators used in these operations can be adjusted to construct different quantum erasure-correcting codes. A special feature of this scheme is that no measurement is required. To illustrate our scheme, we present an example in which five-qubits of information are protected against the occurrence of two erasures. 
}{}{}

\vspace*{10pt}

\vspace*{3pt}

\vspace*{1pt}\textlineskip    
\section{Introduction}\label{intro}

The use of quantum systems in various applications of computing and information processing is subject to the mitigation of the effects of a phenomenon known as {\it decoherence} which can be seen as a consequence of quantum entanglement between the system and the environment \cite{Shor1995,Bus2007}. One of the implications of decoherence is the occurrence of loss of quantum information. For example, photons are usually lost in transmission line (corresponding to an erasure in the information theoretic language) which represents a significant obstacle to the survival of quantum coherence \cite{Lassen2010}.


Erasure-correcting codes have long been known in classical coding theory, and their quantum counterparts have also been theoretically developed. A special class of quantum erasure-correction code was proposed by Grassl et al. \cite{Gras1997} who consi\-de\-red a situation in which the position of the erroneous (lost) qubits is known. According to classical coding theory, they called this model the {\it quantum erasure channel} (QEC). Some physical scenarios to determine the position of an error, such as spontaneous emission, have been given in the literature \cite{Gras1997}.

In general, alteration of information is not {\it a priori} obvious for the observer, which should encode the information in a special way to detect such change. One way that can be explored to perform such an encoding is using Greenberger-Horne-Zeilinger (GHZ) states. The {\it GHZ state} (also called {\it cat} state) was introduced by Daniel M. Greenberger, Michael A. Horne and Anton Zeilinger \cite{GHZ:89} as a new way of proving Bell's Theorem \cite{Bell:64}. 

In past few years, GHZ states have been extensively  studied by many researchers. They play an important role in quantum information processing and communication \cite{Bouwm1999,Chen2008}. As the most frequently used multiparty entangled state, the GHZ state has appeared in applications such as nonlocality \cite{Mermin:argGHZ}, multiparty quantum communication \cite{Yang2004,Buhrman:qcomcomplexity}, and multiparty cryptography \cite{Brassard:qcommunic}.

Given that there are few codes that addressed the retrieval of information upon the occurrence of erasure and also by the importance that this change represents for various scenarios in quantum computation and communication, Yang et al. \cite{Yang:schemGHZ3q} presented a code which protects three-qubits of  information against one erasure using GHZ states. By developing a genera\-li\-zation of this code, making use of a single block of redundancy to handle any number $k\geq 3$ qubits, we realized that simply increasing the number of qubits can only protect $k$-qubits of information against the occurrence of only one erasure \cite{Santos:csIT2010}.

From the point of view of some practical applications, such as in Josephson junctions \cite{Fazio1999}, neutral atoms in optical lattices \cite{Vala2005}, and, most notoriously, in single photons that can be lost during processing or due to inefficient photon sources and detectors \cite{NLM2001,WaBa2007}, it should be noted that  the occurrence of erasure is hardly restricted to only one qubit. 


Lassen et al. \cite{Lassen2010} presented a first experimental realization of an apparatus capable of protecting against the occurrence of quantum erasures. However, this apparatus has been developed for quantum continuous-variable systems which have canonical coordinates corres\-ponding to position and momentum, for instance, as in the {\it quantum harmonic oscillator}. These observables do not have a discrete set of eigenvalues, but a continuous spectrum of them. Hence, the term {\it continuous-variable systems} has been coined to describe this type of situation \cite{Eisert2007}.  One limitation of working with these systems is that one does not have a complete control over failures that occur in operations. This difficulty arises because the underlying Hilbert space is infinite dimensional \cite{Plenio2006}.

In this paper, we will characterize a scheme that has a discrete set of eigenvalues to protect the information against the occurrence of multiple erasures by improving the code given by Yang et al. \cite{Yang:schemGHZ3q}. Our technique allows to protect $k$-qubit ($k\geq 3$) of information against the occurrence of $t = \lfloor k / 2 \rfloor$ erasures. In this scheme ($t + 1$) redundant blocks are used and we restrict to the case that each erasure must occur in distinct blocks (sent through different channels). The detection of the occurrence of erasures in different channels is already commonly used in practical experiments \cite{Lassen2010,Niset2008}.

We stress that a special feature of this scheme is that no measurement is required, since information about the erasures is provided naturally by the system (e.g., spontaneous emission) and also because the restoring operation consists of unitary operators. In addition, codes constructed by this scheme can work when the interaction with the environment causes a leakage out of the qubit space. We will show how the present scheme work through the encoding and restoring operations.



This paper is organized as follows. Section \ref{overview}  introduces a scheme for protecting quantum information against multiple erasures using the GHZ states. Section \ref{schemerasures} shows the formulation of the encoding operation and restoring operation that enable the protection of information against the occurrence of  multiple erasures. To illustrate the proposed scheme, in Section \ref{examplescheme} we show an example where five-qubits of information are protected against the occurrence of two erasures. Finally, in Section \ref{conclus}, we present our concluding remarks.

\section{General idea of the proposed scheme}\label{overview}


Our idea is to improve the code given by Yang et al. \cite{Yang:schemGHZ3q} aiming at developing a scheme that has a discrete set of eigenvalues to protect the information against the occurrence of multiple erasures. One possibility to construct this code could be based on the growth of the number of redundancy blocks. But, in consequence, the following questions arise: 

\begin{itemize}
\item How to increase the number of blocks? 
\item What would be the length of each block? 
\end{itemize}

In order to answer these questions, we performed an analysis on the encoding, decoding and recovery operations to verify how the amount of redundancy blocks could be increased to enable protection against the occurrence of multiple erasures. In this analysis, we verified that to protect $k\geq 3$ information qubits against the occurrence of $t = \lfloor k / 2 \rfloor$ erasures, it was necessary to use $t+1$ redundant blocks. 

Cerf and Cleve \cite{Cerf:1997} demonstrated that quantum information can be distributed over many qubits through a suitable encoding and subsequently recovered after partial alteration, without violating the no-cloning theorem. In that paper, they showed that, for an arbitrary entanglement between the logical words and a reference system to be preserved, the quantum mutual information between this reference and any interacting part of the codewords must be vanishing prior to decoherence.

Since we want the proposed scheme to have a reference system that is statistically independent of any arbitrarily chosen part, among those who will interact with the environment, then we can consider the situation in which each of the $(t +1)$ blocks of $k \geq 3$ qubits ($t = \lfloor k / 2 \rfloor$) is sent by an independent  channel in a way that the reference system is obtained via blocks of qubits that remain undamaged (i.e. without the occurrence of erasure) after passing through the QEC, which we call {\it undamaged blocks}. Thus, to protect the information against $t$ erasures  we use $k$ qubits in each channel.

Although we were dealing with a special case in which a single party cannot obtain any information about the state as a whole, the purpose here is to present a concrete scheme to protect $k$ qubits information against the occurrence of $t$ erasures.

We will now briefly describe the three steps that comprise the proposed scheme to protect the information against the occurrence of multiple erasures:

\begin{description}
\item a)  \ We prepare an arbitrary state $\ket{\psi}_{k}$ of $k \geq 3$ qubits to be transmitted, as well as the $t$ blocks of $k$  ancillary qubits each (all initially in state $\vert 0^{\otimes k} \rangle$), where $t= \lfloor k / 2 \rfloor$. After that, it is applied the encoding operator $U_{enc}$ to the product of $\ket{\psi}_{k}$ state with the $t$ blocks of auxiliary qubits, in such a way to transform each one of the $2^{k}$ basis states of the $\ket{\psi}_{k}$ as a product of $(t +1)$ identical blocks of GHZ states of $k$ qubits each; 
\item b)  \ Each one of the $(t +1)$ blocks of encoded state is sent through $(t +1)$ independent channels, which may suffer up to $t$ erasures (recall that each erasure must occur in distinct blocks);
\item c)  \ The corrupted state is recovered through the restoring operation. This operation makes use of another block of $k$ ancillary qubits, of the decoding operator $U_{dec} $, and of the recovery operator $U_{rec}$. If in a given channel one erasure occur, the block of qubits inherent in that channel is handled by the $U_{rec}$ operator, otherwise it will be worked by the $U_{dec}$ operator. After the application of these operators, we obtain $| B |$ GHZ states,\footnote{$B\subset D$, $D=\{ 0, \dots , t\}$, is the set of blocks in which erasures were detected.} \ all in the form $1/\sqrt{2} (\vert 0^{\otimes k} \rangle \pm \vert 1^{\otimes k} \rangle)$, which is called {\it canonical GHZ state}, and also $(t+1- |B|)$ blocks of $\vert 0^{\otimes k} \rangle$ states. With this, we can then separate the $\ket{\psi}_{k}$ state via the block of index $(t +1)$, now free of erasures.
\end{description}

The next section will show the formulation of encoding and restoring operations to the realization of the proposed scheme.

\section{Encoding and Restoring Operations}\label{schemerasures} 

In this section we show the formulation of the encoding and restoring operations that enable the protection of information against the occurrence of  $t = \lfloor k / 2 \rfloor$ quantum erasures. 

We will use the following notation: 

\begin{itemize}
\item $\overbrace{\ket{0} \otimes \ldots \otimes \ket{0}}^{k} =  \vert \overbrace{ 0 \ldots 0 }^{k} \rangle =\vert 0^{\otimes k}  \rangle$;
\item $\bigotimes_{d=1}^{n} \vert 0^{\otimes k} \rangle_{(d)}$ to denote the tensor product sequence $\vert 0^{\otimes k} \rangle_{(1)} \otimes \ldots \otimes \vert 0^{\otimes k} \rangle_{(n)}$;
\item $m(d)$ stands for the position $m$ of a qubit  in the block of index $(d)$;
\item $|V|$ denotes the cardinality of $V$.
\end{itemize}

Let $\ket{\psi}$ be an arbitrary state of $k\geq 3$ qubits. We can encode the $\ket{\psi}$ state into

%

\begin{equation}\label{GHZstates}
\vert \psi \rangle_{GHZ} = \frac{1}{\sqrt{2^{t+1}}} \sum_{i=0}^{2^{k}-1} \lambda_{i} \bigotimes_{d=0}^{t} \bigg[ \Big\vert u_{1(d)}^{(i)} u_{2(d)}^{(i)} \ldots u_{k(d)}^{(i)} \Big\rangle + (-1)^{i}  \Big\vert \hat{u}_{1(d)}^{(i)} \hat{u}_{2(d)}^{(i)} \ldots \hat{u}_{k(d)}^{(i)} \Big\rangle \bigg]  ,
\end{equation} 
where $\sum_{i=0}^{2^{k}-1} |\lambda_{i}|^{2}=1$ and $(d)$ refers to blocks of k qubits as follows: the block of index $(0)$ corresponds to the first $k$ qubits (the {\it message}), while the block of indices $(1)$ to $(t)$ correspond to blocks of $k$ ancillary qubits each, respectively. Here, $\vert  u_{m(d)}^{(i)} \rangle$ and $\vert  \hat{u}_{m(d)}^{(i)} \rangle$ represent two orthogonal states of the qubit in the position $m(d)$, where $\hat{u}_{m(d)}^{(i)} = 1-u_{m(d)}^{(i)}$ and $u_{m(d)}^{(i)} \in \{ 0, 1 \}$.

Since the $\ket{\psi}_{GHZ}$ state is composed by a product of $(t +1)$ blocks of identical $k$-qubit GHZ states each, it is straightforward to show that for the encoded state (\ref{GHZstates}), the density operator of each qubit is given by $\frac{1}{2} (\vert 0 \rangle \langle 0 \vert + \vert 1 \rangle \langle 1 \vert)$. This result means that the $k$-qubit quantum information, originally carried by the $k$  message qubits, is distributed over each qubit after encoding the $\ket{\psi}$ state  into the $\ket{\psi}_{GHZ}$ state.

With the completion of encoding, given in (\ref{GHZstates}), we will obtain $(t +1)$ redundant blocks, all in GHZ basis. As a result of redundancy, the $k$-qubit quantum state originally encoded may be recovered by a restoring operation when it suffer erasures.


The encoding operation given in (\ref{GHZstates}) can be easily build using Hadamard gates and Controled-NOT (CNOT) gates, according to the following steps:

\begin{enumerate}\label{stepcod}
\item Each basis states of $\ket{\psi}$ is identically prepared in the $t$ blocks of ancillary qubits via a unitary operator $U_{red}$, that makes use of CNOT operations. As a result, the state is immersed in a $2^{k (t +1)}$-dimensional space.
\item In the $k$-th qubit of each block we apply the Hadamard transform, and as a result we have that the $k$-th qubit of each block will be now an addition or a subtraction, depending on whether the $k$-th qubit is in state $\ket{0}$ or in state $\ket{1}$.
\item Finally, it is used a unitary operator $U_{GHZ} $, consisting of CNOT operations, which acts on each block in such a way to make qubits of the second term of the addition (or subtraction) is the complement of qubits of the first term, similarly to the expression (\ref{GHZstates}).
\end{enumerate}


After the three steps described, we get a state composed of a product of $(t +1)$ identical blocks in the GHZ basis of $k$ qubits each. It is important to note that, after completing the encoding, no amplitude has been changed.

The workings of the step $1$ of encoding is described by the following lemma. To do so, we show that with an arbitrary state $\ket{\psi}$ of $k$ qubits and doing it with the tensor product of $t$ blocks of $k$ ancillary qubits each (where all qubits are initially in $\ket{0}$ state), we obtain a state immersed in the $2^{k (t +1)}$-dimensional space in order to make sure that it is composed of $t+1$ identical blocks in the computational basis.

\begin{lemma}\label{opredund}
Let $\ket{ \psi }$ be an arbitrary state of $k$-qubits ($k \geq 3$) in the computational basis and $t=\lfloor k/2 \rfloor$ ancillary blocks of $k$ qubits each, all initially in $\vert 0 \rangle$ state. Then, the unitary linear operator $U_{red}$ encodes the product of the $\ket{ \psi }$ state with $t$ auxiliary blocks in such a way that the result is the product of $t+1$ identical blocks to basis states of $\ket{ \psi }$ (immersion in a $2^{k(t+1)}$-dimensional space), where

\begin{eqnarray}\label{cnotoperat}
U_{red} &=& \prod_{d=1}^{t} \left( \prod_{i=1}^{k} C_{i(0),i(d)} \right) 
\end{eqnarray}\\
and $C_{x,y}$ is a CNOT operation acting on the qubit $y$ (target bit) controled by the state of qubit $x$ (control bit).
\end{lemma}

\noindent{\bf Proof:} An arbitrary state $\ket{\psi}$ of $k$ qbits ($k \geq 3$) can be described by binary decomposition, as follows:

\begin{equation}\label{decombin}
\ket{\psi} = \lambda_{0} \ket{0_{1}0_{2}\cdots 0_{k}} + \lambda_{1} \ket{0_{1}0_{2}\cdots 1_{k}} + \cdots + \lambda_{2^{k}-1} \ket{1_{1}1_{2}\cdots 1_{k}}.
\end{equation}\\
where $\sum_{i=0}^{2^{k}-1} |\lambda_{i}|^{2}=1$.

The tensor product of $\ket{\psi}$ with $d=\{1, \ldots , t \}$ blocks of $k$ ancillary qubits, all in the state $\vert 0 \rangle$, is given as follows

\begin{flushleft}
\begin{eqnarray}\label{prov1a}
 \ket{\psi}_{(0)} \bigotimes_{d=1}^{t} \vert 0 \rangle_{(d)}^{\otimes k} & = & \Big( \lambda_{0} \ket{0_{1}0_{2}\cdots 0_{k}}_{(0)} + \lambda_{1} \ket{0_{1}0_{2}\cdots 1_{k}}_{(0)} + \cdots  \nonumber \\
& & + \lambda_{2^{k}-1} \ket{1_{1}1_{2} \cdots 1_{k}}_{(0)} \Big) \otimes \Big( \ket{0_{1}0_{2}\cdots 0_{k}}_{(1)} \otimes \ldots \nonumber \\
& & \otimes \ket{0_{1}0_{2}\cdots 0_{k}}_{(t)} \Big) \nonumber \\
& = & \lambda_{0} \Big( \ket{0_{1}0_{2}\cdots 0_{k}}_{(0)} \otimes \ket{0_{1}0_{2}\cdots 0_{k}}_{(1)} \otimes \ldots  \nonumber \\
& & \otimes \ket{0_{1}0_{2}\cdots 0_{k}}_{(t)} \Big) + \lambda_{1} \Big( \ket{0_{1}0_{2}\cdots 1_{k}}_{(0)} \otimes \ket{0_{1}0_{2}\cdots 0_{k}}_{(1)} \nonumber \\
& &  \otimes \ldots \otimes \ket{0_{1}0_{2}\cdots 0_{k}}_{(t)} \Big) + \cdots + \lambda_{2^{k}-1} \Big( \ket{1_{1}1_{2}\cdots 1_{k}}_{(0)} \nonumber \\
& &  \otimes \ket{0_{1}0_{2}\cdots 0_{k}}_{(1)} \otimes \ldots \otimes \ket{0_{1}0_{2}\cdots 0_{k}}_{(t)} \Big) .
\end{eqnarray}
\end{flushleft}

Since $U_{red} $ is a linear operator, its application in (\ref{prov1a}), results:

\begin{eqnarray}\label{prov1bb}
\ket{\psi}' & = & U_{red} \Big( \ket{\psi}_{(0)} \bigotimes_{d=1}^{t} \vert 0 \rangle_{(d)}^{\otimes k} \Big) \nonumber \\
 & = & U_{red} \Big[ \lambda_{0} \big( \ket{0_{1}0_{2}\cdots 0_{k}}_{(0)} \otimes \ket{0_{1}0_{2}\cdots 0_{k}}_{(1)} \otimes \ldots  \otimes \ket{0_{1}0_{2}\cdots 0_{k}}_{(t)} \big) \Big]  \nonumber \\
& & +  U_{red} \Big[  \lambda_{1} \big( \ket{0_{1}0_{2}\cdots 1_{k}}_{(0)} \otimes \ket{0_{1}0_{2}\cdots 0_{k}}_{(1)} \otimes \ldots \otimes \ket{0_{1}0_{2}\cdots 0_{k}}_{(t)} \big) \Big] \nonumber \\
& & + \cdots +  U_{red} \Big[  \lambda_{2^{k}-1} \big( \ket{1_{1}1_{2}\cdots 1_{k}}_{(0)} \otimes \ket{0_{1}0_{2}\cdots 0_{k}}_{(1)} \otimes \ldots \nonumber \\
& &  \otimes \ket{0_{1}0_{2}\cdots 0_{k}}_{(t)} \big) \Big] . 
\end{eqnarray}\\
We have that

\begin{eqnarray}\label{prov1c}
U_{red} &=& \prod_{d=1}^{t} \left( \prod_{i=1}^{k} C_{i(0),i(d)} \right) \nonumber \\
           & = & \Big( C_{1(0),1(1)} C_{2(0),2(1)} \cdots C_{k(0),k(1)} \Big) \cdots \Big( C_{1(0),1(t)} C_{2(0),2(t)} \cdots C_{k(0),k(t)} \Big) , \nonumber \\
           & &
\end{eqnarray}\\
in which the compositions of the CNOT operations are performed from right to left.

As we can see in (\ref{prov1c}), for each application of the $C_ {x, y}$ operation the position of the control bit, which is always observed in the index block $(0)$, is equal the position of the target bit to be applied in the block index $(d)$, where $d \in \{ 1, \ldots , t \}$, for each $k$ positions.

Performing now the application of (\ref{prov1c}) in (\ref{prov1bb}), we obtain

\begin{eqnarray}\label{prov1d}
\ket{\psi}' & = & \lambda_{0} \Big( \ket{0_{1}0_{2}\cdots 0_{k}}_{(0)} \otimes \ket{0_{1}0_{2}\cdots 0_{k}}_{(1)} \otimes \ldots \otimes \ket{0_{1}0_{2}\cdots 0_{k}}_{(t)} \Big) \nonumber \\
 & & + \lambda_{1} \Big( \ket{0_{1}0_{2}\cdots 1_{k}}_{(0)} \otimes \ket{0_{1}0_{2}\cdots 1_{k}}_{(1)} \otimes \ldots \otimes \ket{0_{1}0_{2}\cdots 1_{k}}_{(t)}  \nonumber \Big) + \cdots \\
 & & + \lambda_{2^{k}-1} \Big( \ket{1_{1}1_{2}\cdots 1_{k}}_{(0)} \otimes \ket{1_{1}1_{2}\cdots 1_{k}}_{(1)} \otimes \ldots \otimes \ket{1_{1}1_{2}\cdots 1_{k}}_{(t)} \Big) . \nonumber \\
 & &
\end{eqnarray}

Therefore, after applying the $U_{red}$ operator  to the product $\Big( \ket{\psi}_{(0)} \bigotimes_{d=1}^{t} \vert 0 \rangle_{(d)}^{\otimes k} \Big)$, we obtain a state composed of $t + 1$ blocks identical to the basis states of $\ket{\psi}$, where $t=\lfloor k / 2 \rfloor$. 
\begin{flushright}
$\square$
\end{flushright}


To perform the step $2$ of encoding (p. \pageref{stepcod}), we apply the Hadamard transform to the $k$-th qubit of each one of the $(t +1)$ blocks of (\ref{prov1d}), i.e., $H_{k} \ket{\psi}'$. Thus, we get (normalization factors are omitted):

 \begin{eqnarray}\label{operHad}
\ket{\psi} '' = H_{k} \ket{\psi}' & = & \lambda_{0} \Big[ \big( \ket{0_{1}0_{2}\cdots 0_{k-1}0_{k}} + \ket{0_{1}0_{2}\cdots 0_{k-1}1_{k}}\big)_{(0)} \nonumber \\
 & & \otimes \big( \ket{0_{1}0_{2}\cdots 0_{k-1}0_{k}} + \ket{0_{1}0_{2}\cdots 0_{k-1}1_{k}}\big)_{(1)}  \nonumber \\
& & \otimes \ldots \otimes \big( \ket{0_{1}0_{2}\cdots 0_{k-1}0_{k}} + \ket{0_{1}0_{2}\cdots 0_{k-1}1_{k}}\big)_{(t)} \Big] \nonumber \\
& & + \lambda_{1} \Big[ \big( \ket{0_{1}0_{2}\cdots 0_{k-1}0_{k}} - \ket{0_{1}0_{2}\cdots 0_{k-1}1_{k}} \big)_{(0)} \nonumber \\
& & \otimes \big( \ket{0_{1}0_{2}\cdots 0_{k-1}0_{k}} - \ket{0_{1}0_{2}\cdots 0_{k-1}1_{k}} \big)_{(1)}  \nonumber \\
& & \otimes \ldots \otimes \big( \ket{0_{1}0_{2}\cdots 0_{k-1}0_{k}} - \ket{0_{1}0_{2}\cdots 0_{k-1}1_{k}} \big)_{(t)} \Big] + \cdots \nonumber \\
& & + \lambda_{2^{k}-1} \Big[ \big( \ket{1_{1}1_{2}\cdots 1_{k-1}0_{k}} -  \ket{1_{1}1_{2}\cdots 1_{k-1}1_{k}}  \big)_{(0)} \nonumber \\
& & \otimes \big( \ket{1_{1}1_{2}\cdots 1_{k-1}0_{k}} -  \ket{1_{1}1_{2}\cdots 1_{k-1}1_{k}}  \big)_{(1)} \nonumber \\
& & \otimes \ldots \otimes \big( \ket{1_{1}1_{2}\cdots 1_{k-1}0_{k}} -  \ket{1_{1}1_{2}\cdots 1_{k-1}1_{k}} \big)_{(t)} \Big] . 
\end{eqnarray}

It will be shown, in the following lemma, that the step $3$ of encoding (p. \pageref{stepcod}) can be performed by a unitary operation in $\ket{\psi}''$ such that, in the second term, each addition/subtraction in each block of $k$ qubits is the complement of the first term.

\begin{lemma}\label{operaghz}
Let $\ket{\psi}''$ a state composed of $t +1$ identical blocks of $k$ qubits each, as described in (\ref{operHad}), where $k \geq 3$ and $t = \lfloor k / 2 \rfloor$. Then the unitary linear operator

\begin{eqnarray}\label{ghzoperat}
U_{GHZ} &=& \prod_{d=0}^{t} \Bigg( \prod_{i=1}^{k-1} C_{k(d),i(d)} \Bigg) 
\end{eqnarray}\\
encodes the $\ket{\psi}''$ state such that the second term of each addition/subtraction, in each block of $k$ qubits, is the complement of the first term.
\end{lemma}

\noindent{\bf Proof:} Given $\ket{\psi}''$ state as described in (\ref{operHad}) and considering $U_{GHZ}$ as a unitary linear operator, then applying $U_{GHZ}$ in (\ref{operHad}) results in:

\begin{eqnarray}\label{GHZpsi}
\ket{\psi}_{GHZ} & = & U_{GHZ} \big( \ket{\psi} '' \big) \nonumber \\
 & = & \lambda_{0}  \bigg\{ U_{GHZ}  \Big[ \big( \ket{0_{1}0_{2}\cdots 0_{k-1}0_{k}} + \ket{0_{1}0_{2}\cdots 0_{k-1}1_{k}}\big)_{(0)} \nonumber \\
 & & \otimes \big( \ket{0_{1}0_{2}\cdots 0_{k-1}0_{k}} + \ket{0_{1}0_{2}\cdots 0_{k-1}1_{k}}\big)_{(1)}  \nonumber \\
& & \otimes \ldots \otimes \big( \ket{0_{1}0_{2}\cdots 0_{k-1}0_{k}} + \ket{0_{1}0_{2}\cdots 0_{k-1}1_{k}}\big)_{(t)} \Big] \bigg\}  \nonumber \\
& & + \lambda_{1} \bigg\{ U_{GHZ} \Big[ \big( \ket{0_{1}0_{2}\cdots 0_{k-1}0_{k}} - \ket{0_{1}0_{2}\cdots 0_{k-1}1_{k}} \big)_{(0)} \nonumber \\
& & \otimes \big( \ket{0_{1}0_{2}\cdots 0_{k-1}0_{k}} - \ket{0_{1}0_{2}\cdots 0_{k-1}1_{k}} \big)_{(1)}  \nonumber \\
& & \otimes \ldots \otimes \big( \ket{0_{1}0_{2}\cdots 0_{k-1}0_{k}} - \ket{0_{1}0_{2}\cdots 0_{k-1}1_{k}} \big)_{(t)} \Big] \bigg\} + \cdots  \nonumber \\
& & + \lambda_{2^{k}-1} \bigg\{ U_{GHZ} \Big[ \big( \ket{1_{1}1_{2}\cdots 1_{k-1}0_{k}} -  \ket{1_{1}1_{2}\cdots 1_{k-1}1_{k}}  \big)_{(0)} \nonumber \\
& & \otimes \big( \ket{1_{1}1_{2}\cdots 1_{k-1}0_{k}} -  \ket{1_{1}1_{2}\cdots 1_{k-1}1_{k}}  \big)_{(1)} \nonumber \\
& & \otimes \ldots \otimes \big( \ket{1_{1}1_{2}\cdots 1_{k-1}0_{k}} -  \ket{1_{1}1_{2}\cdots 1_{k-1}1_{k}} \big)_{(t)} \Big] \bigg\} . 
\end{eqnarray}\\
Note that:

\begin{eqnarray}\label{ghzop2}
U_{GHZ} & = & \prod_{d=0}^{t} \Bigg( \prod_{i=1}^{k-1} C_{k(d),i(d)} \Bigg) \nonumber \\
            & = & \Big( C_{k(0),1(0)}C_{k(0),2(0)}\cdots C_{k(0),k-1(0)} \Big) \Big( C_{k(1),1(1)}C_{k(1),2(1)}\cdots C_{k(1),k-1(1)} \Big) \nonumber \\
            &  & \cdots \Big( C_{k(t),1(t)}C_{k(t),2(t)}\cdots C_{k(t),k-1(t)} \Big) .   
\end{eqnarray}

In order to make each addition/subtraction in each block of k qubits in the second term be the complement of the first one, the operator $C_{x, y}$ in (\ref{ghzop2}) must acts on qubits that are in the positions $1$ to $k-1$, for each $t +1$ blocks, considering the qubit of $k$-th position (control bit), as follows: if it is in the $\ket{1}$ state then the qubits $b_{i}$ (where $i = 1, \ldots, k-1$ indicates the position of the qubit) will be changed to $(b_{i} + 1 \ \mbox{mod} \ 2)$; otherwise there will not be changes.

Performing now the application of (\ref{ghzop2}) in (\ref{GHZpsi}), we obtain

\begin{eqnarray}\label{aplghzop2}
\ket{\psi}_{GHZ} & = & U_{GHZ} \big( \ket{\psi} '' \big) \nonumber \\
& = & \lambda_{0} \Big[ \big( \ket{0_{1}0_{2}\cdots 0_{k-1}0_{k}} + \ket{1_{1}1_{2}\cdots 1_{k-1}1_{k}} \big)_{(0)} \nonumber \\
& &  \otimes \big( \ket{0_{1}0_{2}\cdots 0_{k-1}0_{k}} + \ket{1_{1}1_{2}\cdots 1_{k-1}1_{k}} \big)_{(1)}  \nonumber \\
& & \otimes \ldots \otimes \big( \ket{0_{1}0_{2}\cdots 0_{k-1}0_{k}} + \ket{1_{1}1_{2}\cdots 1_{k-1}1_{k}} \big)_{(t)} \Big] \nonumber \\
& & + \lambda_{1} \Big[ \big( \ket{0_{1}0_{2}\cdots 0_{k-1}0_{k}} - \ket{1_{1}1_{2}\cdots 1_{k-1}1_{k}} \big)_{(0)} \nonumber \\
& & \otimes \big( \ket{0_{1}0_{2}\cdots 0_{k-1}0_{k}} - \ket{1_{1}1_{2}\cdots 1_{k-1}1_{k}} \big)_{(1)} \nonumber \\
& & \otimes \ldots \otimes \big( \ket{0_{1}0_{2}\cdots 0_{k-1}0_{k}} - \ket{1_{1}1_{2}\cdots 1_{k-1}1_{k}} \big)_{(t)} \Big] + \cdots \nonumber \\
& &  + \lambda_{2^{k}-1} \Big[ \big( \ket{1_{1}1_{2}\cdots 1_{k-1}0_{k}} -  \ket{0_{1}0_{2}\cdots 0_{k-1}1_{k}}  \big)_{(0)} \nonumber \\
& & \otimes \big( \ket{1_{1}1_{2}\cdots 1_{k-1}0_{k}} -  \ket{0_{1}0_{2}\cdots 0_{k-1}1_{k}}  \big)_{(1)} \nonumber \\
& & \otimes \ldots \otimes \big( \ket{1_{1}1_{2}\cdots 1_{k-1}0_{k}} -  \ket{0_{1}0_{2}\cdots 0_{k-1}1_{k}}  \big)_{(t)} \Big] . 
\end{eqnarray}

Therefore, after applying the operator $U_{GHZ}$ in $\ket{\psi}''$, we obtain a $\ket{\psi} _ {GHZ}$ state such that the second term of each addition/subtraction, in each block of $k$ qubits, is the complement of the first term.
\begin{flushright}
$\square$
\end{flushright}


The following theorem shows the operation that encodes each of the basis states of $\ket{\psi}$, of $k$ qubits, as a product of $(t +1)$ redundant blocks of GHZ states of $k$-qubits each.

\begin{theorem}\label{encapag}
Let $\ket{\psi}$ be an arbitrary state of $k$-qubits ($k \geq 3$) in the computational basis and $t = \lfloor k / 2 \rfloor$ blocks of $k$ ancillary qubits each, all initially in $\vert 0 \rangle$ state. Then, the encoding operation, denoted by $\mathfrak{E}_{GHZ}$, encodes each of the basis states of $\ket{\psi}$ as a product of $(t +1)$ redundant blocks of GHZ states of $k$-qubits each. This encoding operation $\mathfrak{E}_{GHZ}$ is given by

\begin{equation}\label{encoperat}
\mathfrak{E}_{GHZ} =U_{enc} \bigg[ \vert \psi \rangle_{(0)} \bigotimes_{d=1}^{t} \Big( \vert 0^{\otimes k} \rangle_{(d)} \Big) \bigg],
\end{equation}
where

\begin{eqnarray}\label{opencod}
U_{enc} &=& U_{GHZ} \cdot \left( \prod_{d=0}^{t} H_{k(d)} \right) \cdot U_{red} ,
\end{eqnarray} 
and $U_{red}$ as in (\ref{cnotoperat}) and $U_{GHZ}$ as in (\ref{ghzoperat}).
\end{theorem}

\noindent{\bf Proof:} Let $\ket{\psi}$ be an arbitrary state of $k$ qubits ($k \geq 3$) which is described, by binary decomposition, as follows

\begin{eqnarray}\label{decombin2}
\ket{\psi} & = & \lambda_{0} \ket{0_{1}0_{2}\cdots 0_{k-1}0_{k}} + \lambda_{1} \ket{0_{1}0_{2}\cdots 0_{k-1}1_{k}} + \cdots + \lambda_{2^{k}-1} \ket{1_{1}1_{2}\cdots 1_{k-1}1_{k}}.
\end{eqnarray}\\
where $\sum_{i=0}^{2^{k}-1} |\lambda_{i}|^{2}=1$. 

Now we will apply the $U_{enc}$ operator, given in (\ref{opencod}), to $\Big[ \vert \psi \rangle_{(0)} \bigotimes_{d=1}^{t} \big( \vert 0^{\otimes k} \rangle_{(d)} \big) \Big]$.

By Lemma \ref{opredund}, after applying $U_{red}$ to $\Big[ \vert \psi \rangle_{(0)} \bigotimes_{d=1}^{t} \big( \vert 0^{\otimes k} \rangle_{(d)} \big) \Big]$, we have

\begin{eqnarray}\label{provt1a}
\ket{\psi} ' & = & U_{red} \Big[ \vert \psi \rangle_{(0)} \bigotimes_{d=1}^{t} \big( \vert 0^{\otimes k} \rangle_{(d)} \big) \Big] \nonumber \\
 & = & \lambda_{0} \big( \ket{0_{1}0_{2}\cdots 0_{k-1}0_{k}}_{(0)} \otimes \ket{0_{1}0_{2}\cdots 0_{k-1}0_{k}}_{(1)}  \nonumber \\
& & \otimes \ldots \otimes \ket{0_{1}0_{2}\cdots 0_{k-1}0_{k}}_{(t)} \big) \nonumber \\
& & + \lambda_{1} \big( \ket{0_{1}0_{2}\cdots 0_{k-1}1_{k}}_{(0)} \otimes \ket{0_{1}0_{2}\cdots 0_{k-1}1_{k}}_{(1)} \nonumber \\
& & \otimes \ldots \otimes \ket{0_{1}0_{2}\cdots 0_{k-1}1_{k}}_{(t)} \big) + \cdots \nonumber \\
& &  + \lambda_{2^{k}-1} \big( \ket{1_{1}1_{2}\cdots 1_{k-1}1_{k}}_{(0)} \otimes \ket{1_{1}1_{2}\cdots 1_{k-1}1_{k}}_{(1)} \nonumber \\
& & \otimes \ldots \otimes \ket{1_{1}1_{2}\cdots 1_{k-1}1_{k}}_{(t)} \big) . 
\end{eqnarray}

Applying the Hadamard transform of the $k$-th qubit of each $(t +1)$ blocks of $\ket{\psi}'$, we get (normalization factors are omitted):

\begin{eqnarray}\label{provt1b}
\ket{\psi} '' & = & H_{k} \ket{\psi}' \nonumber \\
& = & \lambda_{0} \Big[ \big( \ket{0_{1}0_{2}\cdots 0_{k-1}0_{k}} + \ket{0_{1}0_{2}\cdots 0_{k-1}1_{k}}\big)_{(0)} \nonumber \\
& & \otimes \big( \ket{0_{1}0_{2}\cdots 0_{k-1}0_{k}} + \ket{0_{1}0_{2}\cdots 0_{k-1}1_{k}}\big)_{(1)}  \nonumber \\
& & \otimes \ldots \otimes \big( \ket{0_{1}0_{2}\cdots 0_{k-1}0_{k}} + \ket{0_{1}0_{2}\cdots 0_{k-1}1_{k}}\big)_{(t)} \Big] \nonumber \\
& & + \lambda_{1} \Big[ \big( \ket{0_{1}0_{2}\cdots 0_{k-1}0_{k}} - \ket{0_{1}0_{2}\cdots 0_{k-1}1_{k}} \big)_{(0)} \nonumber \\
& & \otimes \big( \ket{0_{1}0_{2}\cdots 0_{k-1}0_{k}} - \ket{0_{1}0_{2}\cdots 0_{k-1}1_{k}} \big)_{(1)}  \nonumber \\
& & \otimes \ldots \otimes \big( \ket{0_{1}0_{2}\cdots 0_{k-1}0_{k}} - \ket{0_{1}0_{2}\cdots 0_{k-1}1_{k}} \big)_{(t)} \Big] + \cdots \nonumber \\
& &  + \lambda_{2^{k}-1} \Big[ \big( \ket{1_{1}1_{2}\cdots 1_{k-1}0_{k}} -  \ket{1_{1}1_{2}\cdots 1_{k-1}1_{k}}  \big)_{(0)} \nonumber \\
& & \otimes \big( \ket{1_{1}1_{2}\cdots 1_{k-1}0_{k}} -  \ket{1_{1}1_{2}\cdots 1_{k-1}1_{k}}  \big)_{(1)} \nonumber \\
& & \otimes \ldots \otimes \big( \ket{1_{1}1_{2}\cdots 1_{k-1}0_{k}} -  \ket{1_{1}1_{2}\cdots 1_{k-1}1_{k}} \big)_{(t)} \Big] . 
\end{eqnarray}

By Lemma \ref{operaghz}, after applying $U_{GHZ}$ to $\ket{\psi}''$ state, we have

 \begin{eqnarray}\label{provt1c}
\ket{\psi}_{GHZ} & = & U_{GHZ} \ket{\psi}'' \nonumber \\
& = & \lambda_{0} \Big[ \big( \ket{0_{1}0_{2}\cdots 0_{k-1}0_{k}} + \ket{1_{1}1_{2}\cdots 1_{k-1}1_{k}} \big)_{(0)} \nonumber \\
& &  \otimes \big( \ket{0_{1}0_{2}\cdots 0_{k-1}0_{k}} + \ket{1_{1}1_{2}\cdots 1_{k-1}1_{k}} \big)_{(1)}  \nonumber \\
& & \otimes \ldots \otimes \big( \ket{0_{1}0_{2}\cdots 0_{k-1}0_{k}} + \ket{1_{1}1_{2}\cdots 1_{k-1}1_{k}} \big)_{(t)} \Big] \nonumber \\
& & + \lambda_{1} \Big[ \big( \ket{0_{1}0_{2}\cdots 0_{k-1}0_{k}} - \ket{1_{1}1_{2}\cdots 1_{k-1}1_{k}} \big)_{(0)} \nonumber \\
& & \otimes \big( \ket{0_{1}0_{2}\cdots 0_{k-1}0_{k}} - \ket{1_{1}1_{2}\cdots 1_{k-1}1_{k}} \big)_{(1)} \nonumber \\
& & \otimes \ldots \otimes \big( \ket{0_{1}0_{2}\cdots 0_{k-1}0_{k}} - \ket{1_{1}1_{2}\cdots 1_{k-1}1_{k}} \big)_{(t)} \Big] + \cdots \nonumber \\
& &  + \lambda_{2^{k}-1} \Big[ \big( \ket{1_{1}1_{2}\cdots 1_{k-1}0_{k}} -  \ket{0_{1}0_{2}\cdots 0_{k-1}1_{k}}  \big)_{(0)} \nonumber \\
& & \otimes \big( \ket{1_{1}1_{2}\cdots 1_{k-1}0_{k}} -  \ket{0_{1}0_{2}\cdots 0_{k-1}1_{k}}  \big)_{(1)} \nonumber \\
& & \otimes \ldots \otimes \big( \ket{1_{1}1_{2}\cdots 1_{k-1}0_{k}} -  \ket{0_{1}0_{2}\cdots 0_{k-1}1_{k}}  \big)_{(t)} \Big] . 
\end{eqnarray}

The result presented in (\ref{provt1c}) concludes the application of the encoding operation $\mathfrak{E}_{GHZ}$. 

Therefore, the tensor product of the $\ket{\psi}$ state, of $k$-qubits ($k \geq 3$), with $t = \lfloor k / 2 \rfloor$ ancillary blocks of $k$ qubits each (all initially in the $\ket{0}$ state), is encoded by $\mathfrak{E}_{GHZ}$ in such a way to produce a $\ket{\psi}_{GHZ}$ state, which has ($t + 1$) redundant blocks of $k$ qubits each $(k \geq 3)$ in GHZ basis.
\begin{flushright}
$\square$
\end{flushright}

We can certainly figure out situations where it is possible to know where the error occurred (for methods to determine the position of an error, see \cite{Gras1997}). Because $\ket{0}$ and $\ket{1}$ form a basis for a qubit, we need only to know what happens with these two states. In general, the process of decoherence must be

\begin{eqnarray}\label{env_error}
\vert e_{0} \rangle \vert 0 \rangle & \longrightarrow & \vert \epsilon_{0} \rangle \vert 0 \rangle + \vert \epsilon_{1} \rangle \vert 1 \rangle  , \nonumber \\
\vert e_{0} \rangle \vert 1 \rangle & \longrightarrow & \vert \epsilon_{0}^{'} \rangle \vert 0 \rangle + \vert \epsilon_{1}^{'} \rangle \vert 1 \rangle ,
\end{eqnarray} \\
where $\vert \epsilon_{0} \rangle$, $\vert \epsilon_{1} \rangle$, $\vert \epsilon_{0}^{'} \rangle$ and $\vert \epsilon_{1}^{'} \rangle$ are states of the appropriate environment, not necessarily orthogonal or normalized, and $\ket{e_{0}}$ is the initial state of the environment \cite{Yang:schemGHZ3q}.

As will be shown below, during the restoration operation, there is no need to perform any operations on the erroneous qubit. For simplicity, we can rewrite (\ref{env_error}) as

\begin{eqnarray}\label{env_error2}
\vert e_{0} \rangle \vert 0 \rangle & \longrightarrow & \vert \overline{0} \rangle , \nonumber \\
\vert e_{0} \rangle \vert 1 \rangle & \longrightarrow & \vert \overline{1} \rangle ,
\end{eqnarray} \\  
where  the environment states $\vert \epsilon_{0} \rangle$, $\vert \epsilon_{1} \rangle$, $\vert \epsilon_{0}^{'} \rangle$ and $\vert \epsilon_{1}^{'} \rangle$ in (\ref{env_error}) are included in $\vert \overline{0} \rangle$ end $\vert \overline{1} \rangle$. We assume that any erasure only occurs after the entangled state has been generated. For reference, the $\vert \psi \rangle_{GHZ}$ state after occurrence of an erasure will be represented by $\vert \overline{\psi} \rangle_{GHZ}$. We also admit that at most $t = \lfloor k / 2 \rfloor$ erasures can occur and that they are in distinct blocks. Taking these considerations into account, to restore the state that was originally protected against the occurrence of $t$ erasures, will use the following types of operators:

\begin{itemize}
\item {\it Decoding operator} that act in blocks in which no erasure was detected;
\item {\it Recovery operator},  one for each block in which there were detected erasures.
\end{itemize}

To extract the original state free of erasures, we apply first a unitary transformation on blocks of qubits in which, passing through the QEC, erasures were not detected (which we call {\it undamaged blocks}). This transformation is considered as a partial decoding operator (since the blocks that have undergone erasure are not involved in the decoding operator). To prevent the no-cloning theorem vio\-la\-tion and to facilitate the use of a reference block in the recovery operator, this unitary transformation makes use of a new block of $k$ ancillary qubits (all initially in $\vert 0 \rangle$ state).

This decoding operator, denoted by $U_{dec}$, behaves as follows:

\begin{enumerate}
\item Performs a transformation, from the GHZ basis to the computational basis, in undamaged blocks;
\item The undamaged blocks are identically prepared in the block of index $(t +1)$, consisting of $k$ ancillary qubits, that are initially in the $\ket{0}$ state;
\item Transforms each one of the $k$ qubits of the undamaged blocks into the $\vert 0 \rangle$ state.
\end{enumerate}


The form of this $U_{dec}$ operator is given in the following lemma.

\begin{lemma}\label{opdecghz}
Let $\vert \overline{\psi} \rangle_{GHZ}$ be a state composed of $(t +1)$ identical blocks in GHZ basis of $ k $-qubits each ($ k \geq $ 3), which may have suffered up to $t = \lfloor k / 2 \rfloor$ erasures after passing through the QEC; and let $B \subset D \ (D = \{ 0, \ldots, t \})$ be the set of indices that identify the blocks where erasures were detected. If we apply the unitary linear operator

\begin{equation}\label{opdecod1}
U_{dec} = \prod_{{d=0} \atop { (d \notin B)}}^{t} \Bigg(  \prod_{i=1}^{k} C_{i(t+1),i(d)} \Bigg) \cdot \prod_{{d=0} \atop { (d \notin B)}}^{t} \Bigg( \prod_{i=1}^{k} C_{i(d), i(t+1)} H_{k(d)} \prod_{i=1}^{k-1} C_{k(d),i(d)} \Bigg) , 
\end{equation} 
to the tensor product

$$
\vert \overline{\psi} \rangle_{GHZ} \otimes \vert 0^{\otimes k} \rangle_{(t+1)} ,
$$ 
then following steps take place: $(i)$ all the undamaged blocks of $\vert \overline{\psi} \rangle_{GHZ}$ are transformed from the GHZ basis into the computational basis; $(ii)$ these undamaged blocks are identically prepared in the block of index $(t +1)$; and, lastly, $(iii)$ the qubits of the undamaged blocks are transformed into the $\ket{0}$ state.
\end{lemma}

\noindent{\bf Proof:} Let $\ket{\overline{\psi} }_{GHZ}$ be the state obtained after the $\ket{\psi}_{GHZ}$ state which has $(t+1)$ redundant blocks of $k$ qubits each $(k \geq 3)$ in GHZ basis, passed through the QEC and occurred $t = \lfloor k / 2 \rfloor$ erasures (in that each erasure must occur in distinct blocks).

Since $U_{dec}$ will act only in the undamaged blocks, it is interesting to see its application in two cases:

\begin{description}
\item[1.] When only one block is undamaged ($t$ erasures occured).
\item[2.] When two or more blocks are undamaged. 
\end{description}

For these two cases, we shall show that  the operator (\ref{opdecod1}): (a) will identically prepare, in the block of index $(t +1)$, all the undamaged blocks; and (b) will  transform the $k$ qubits of the undamaged blocks into the $\ket{0}$ state.

\noindent \underline{\bf Case 1:} The $\vert \overline{\psi} \rangle_{GHZ}$ state has $(t +1)$ blocks and suffered $t$ erasures. We consider the case where just one of these blocks is undamaged, i.e., there was the occurrence of these erasures in $t$ different blocks (one erasure in each block). We will establish, without loss of generality, that these erasures occurred in the index blocks of $(0)$ to $(t-1)$, leaving undamaged the block of index ($t$).

Note that for $U_{dec}$, the position where the erasure occurred is not important. However, for purposes of representation, we will assume that  it has been in the position $a$, where $0 \leq a < k$. So, the $\vert \overline{\psi} \rangle_{GHZ}$ state for this case has the following form:

\begin{eqnarray}\label{case1estado}
\vert e_{0} \rangle \otimes  \vert \psi \rangle_{GHZ} \rightarrow \vert \overline{\psi} \rangle_{GHZ} & = & \lambda_{0} \vert \overline{0} \rangle_{L} + \lambda_{1} \vert \overline{1} \rangle_{L} +  \ldots + \lambda_{2^{k}-2} \vert \overline{2^{k}-2} \rangle_{L} + \lambda_{2^{k}-1} \vert \overline{2^{k}-1} \rangle_{L}, \nonumber \\
&  &  
\end{eqnarray}\\
where the logical states  are given as follows, considering that the dash on top represents the position where the erasure occurred and also that the possible phase changes are denoted:

\begin{eqnarray}\label{case1error}
\vert \overline{0} \rangle_{L} & = & \bigg[ \Big( \vert  \ldots \overline{0}_{a} \ldots 0_{k} \rangle \pm \vert  \ldots \overline{1}_{a} \ldots 1_{k} \rangle \Big)_{(0)} \nonumber \\
& & \otimes \ldots \otimes \Big( \vert  \ldots \overline{0}_{a} \ldots 0_{k} \rangle \pm \vert  \ldots \overline{1}_{a} \ldots 1_{k} \rangle \Big)_{(t-1)} \nonumber \\
& & \otimes \Big( \vert \ldots 0_{a} \ldots0_{k} \rangle + \vert \ldots 1_{a} \ldots 1_{k} \rangle \Big)_{(t)} \bigg] , \nonumber \\
\vert \overline{1} \rangle_{L} & = & \bigg[ \Big( \vert  \ldots \overline{0}_{a} \ldots 0_{k} \rangle \mp \vert  \ldots \overline{1}_{a} \ldots 1_{k} \rangle \Big)_{(0)} \nonumber \\
& & \otimes  \ldots \otimes \Big( \vert  \ldots \overline{0}_{a} \ldots 0_{k} \rangle \mp \vert  \ldots \overline{1}_{a} \ldots 1_{k} \rangle \Big)_{(t-1)} \nonumber \\
& & \otimes \Big( \vert \ldots 0_{a} \ldots 0_{k} \rangle - \vert  \ldots 1_{a} \ldots 1_{k} \rangle \Big)_{(t)} \bigg] , \cdots , \nonumber \\
\vert \overline{2^{k}-2} \rangle_{L} & = & \bigg[ \Big( \pm \vert  \ldots \overline{1}_{a} \ldots 0_{k} \rangle + \vert  \ldots \overline{0}_{a} \ldots 1_{k} \rangle \Big)_{(0)} \nonumber \\
& & \otimes \ldots \otimes \Big( \pm \vert  \ldots \overline{1}_{a} \ldots  0_{k} \rangle + \vert  \ldots \overline{0}_{a} \ldots 1_{k} \rangle \Big)_{(t-1)} \nonumber \\
& & \otimes \Big( \vert \ldots 1_{a} \ldots 0_{k} \rangle + \vert  \ldots 0_{a} \ldots 1_{k} \rangle \Big)_{(t)} \bigg] , \nonumber \\
\vert \overline{2^{k}-1} \rangle_{L} & = & \bigg[ \Big( \pm \vert  \ldots \overline{1}_{a} \ldots 0_{k} \rangle - \vert  \ldots \overline{0}_{a} \ldots 1_{k} \rangle \Big)_{(0)} \nonumber \\
& & \otimes \ldots \otimes \Big( \pm \vert  \ldots \overline{1}_{a} \ldots 0_{k} \rangle - \vert  \ldots \overline{0}_{a} \ldots  1_{k} \rangle \Big)_{(t-1)}  \nonumber \\
& & \otimes \Big( \vert \ldots 1_{a} \ldots 0_{k} \rangle - \vert  \ldots 0_{a} \ldots 1_{k} \rangle \Big)_{(t)} \bigg] .
\end{eqnarray}

Since $U_{dec}$ is only applied to the undamaged blocks, this means that, for the case in question, it is applied only to the block of index $(t)$. Therefore, it will be as follows:

\begin{eqnarray}\label{dec_errocase1}
 U_{dec} & = &  \Bigg( \prod_{i=1}^{k} C_{i(t+1),i(t)} \Bigg) \Bigg( \prod_{i=1}^{k} C_{i(t), i(t+1)} \Bigg) H_{k(t)} \Bigg( \prod_{i=1}^{k-1} C_{k(t),i(t)} \Bigg) \nonumber \\
             & = &  \Big( C_{1(t+1), 1(t)} \cdots C_{k(t+1), k(t)} \Big) \Big( C_{1(t), 1(t+1)} \cdots C_{k(t), k(t+1)} \Big) \nonumber \\
              &   &  H_{k(t)} \Big( C_{k(t),1(t)} \cdots C_{k(t),[k-1](t)} \Big) .
\end{eqnarray}

Applying the operator  (\ref{dec_errocase1}) to the product $\Big( \vert \overline{\psi} \rangle_{GHZ} \otimes \vert 0^{\otimes k} \rangle_{(t+1)} \Big)$, we obtain

\begin{eqnarray}\label{case1dec}
\vert \overline{0} \rangle_{L} & = & \bigg[ \Big( \vert  \ldots \overline{0}_{a} \ldots 0_{k} \rangle \pm \vert  \ldots \overline{1}_{a} \ldots 1_{k} \rangle \Big)_{(0)} \nonumber \\
& & \otimes \ldots \otimes \Big( \vert  \ldots \overline{0}_{a} \ldots 0_{k} \rangle \pm \vert  \ldots \overline{1}_{a} \ldots 1_{k} \rangle \Big)_{(t-1)} \nonumber \\
& & \otimes \Big( \vert \ldots 0_{a} \ldots0_{k} \rangle \Big)_{(t)} \otimes \Big( \vert  \ldots 0_{a} \ldots 0_{k} \rangle  \Big)_{(t+1)} \bigg] , \nonumber \\
\vert \overline{1} \rangle_{L} & = & \bigg[ \Big( \vert  \ldots \overline{0}_{a} \ldots 0_{k} \rangle \mp \vert  \ldots \overline{1}_{a} \ldots 1_{k} \rangle \Big)_{(0)} \nonumber \\
& & \otimes  \ldots \otimes \Big( \vert  \ldots \overline{0}_{a} \ldots 0_{k} \rangle \mp \vert  \ldots \overline{1}_{a} \ldots 1_{k} \rangle \Big)_{(t-1)} \nonumber \\
& & \Big( \vert  \ldots 0_{a} \ldots 0_{k} \rangle  \Big)_{(t)} \otimes \Big( \vert \ldots 0_{a} \ldots 1_{k} \rangle \Big)_{(t+1)} \bigg] , \cdots , \nonumber \\
\vert \overline{2^{k}-2} \rangle_{L} & = & \bigg[ \Big( \pm \vert  \ldots \overline{1}_{a} \ldots 0_{k} \rangle + \vert  \ldots \overline{0}_{a} \ldots 1_{k} \rangle \Big)_{(0)} \nonumber \\
& & \otimes \ldots \otimes \Big( \pm \vert  \ldots \overline{1}_{a} \ldots  0_{k} \rangle + \vert  \ldots \overline{0}_{a} \ldots 1_{k} \rangle \Big)_{(t-1)} \nonumber \\
& & \Big( \vert  \ldots 0_{a} \ldots 0_{k} \rangle  \Big)_{(t)} \otimes \Big( \vert \ldots 1_{a} \ldots 0_{k} \rangle \Big)_{(t+1)} \bigg] , \nonumber \\
\vert \overline{2^{k}-1} \rangle_{L} & = & \bigg[ \Big( \pm \vert  \ldots \overline{1}_{a} \ldots 0_{k} \rangle - \vert  \ldots \overline{0}_{a} \ldots 1_{k} \rangle \Big)_{(0)} \nonumber \\
& & \otimes \ldots \otimes \Big( \pm \vert  \ldots \overline{1}_{a} \ldots 0_{k} \rangle - \vert  \ldots \overline{0}_{a} \ldots  1_{k} \rangle \Big)_{(t-1)}  \nonumber \\
& & \Big( \vert  \ldots 0_{a} \ldots 0_{k} \rangle  \Big)_{(t)} \otimes \Big( \vert \ldots 1_{a} \ldots 1_{k} \rangle \Big)_{(t+1)} \bigg] .
\end{eqnarray} 

Note in  (\ref{case1dec}) that, by applying the operator (\ref{dec_errocase1}), the index block $(t)$ left the GHZ basis into the computational basis. After that, this undamaged block was identically prepared in the block of index $(t +1)$ and then had its $k$ qubits transformed to the $\ket{0}$ state. We also emphasize that the blocks of index $(0)$ to $(t-1)$ had no changes after the application of $U_{dec}$, given in (\ref{dec_errocase1}).

\noindent \underline{\bf Case 2:} We will consider that there are $t+1$ undamaged blocks, meaning that no erasure was detected in them. 

Since $U_{dec}$ is only applied to the undamaged blocks, we can explicitly denote it as follows:

\begin{eqnarray}\label{dec_errocase2}
U_{dec} & = & \prod_{d=0}^{t} \Bigg(  \prod_{i=1}^{k} C_{i(t+1),i(d)} \Bigg) \cdot \prod_{d=0 }^{t} \Bigg( \prod_{i=1}^{k} C_{i(d), i(t+1)} H_{k(d)} \prod_{i=1}^{k-1} C_{k(d),i(d)} \Bigg) \nonumber \\
& = & \bigg[ \Big( C_{1(t+1), 1(0)} \cdots C_{k(t+1), k(0)} \Big) \cdots \Big( C_{1(t+1), 1(t)} \cdots C_{k(t+1), k(t)} \Big)  \bigg]   \nonumber \\
              &   & \bigg[ \Big( C_{1(0), 1(t+1)} \cdots C_{k(0), k(t+1)} \Big) H_{k(0)} \Big( C_{k(0),1(0)} \cdots C_{k(0),[k-1](0)} \Big) \cdots \nonumber \\
             &   & \Big( C_{1(t), 1(t+1)} \cdots C_{k(t), k(t+1)} \Big) H_{k(t)}  \Big( C_{k(t),1(t)} \cdots C_{k(t),[k-1](t)} \Big) \bigg] . 
\end{eqnarray}

Now, applying the operator (\ref{dec_errocase2}) in the product $\Big( \vert \overline{\psi} \rangle_{GHZ} \otimes \vert 0^{\otimes k} \rangle_{(t+1)} \Big)$, we obtain

\begin{eqnarray}\label{apldec_errocase2}
\vert \overline{0} \rangle_{L} & = & \bigg[ \Big( \vert 0_{1} \ldots 0_{k-1}0_{k} \rangle \Big)_{(0)} \otimes \ldots \otimes \Big( \vert 0_{1} \ldots 0_{k-1}0_{k} \rangle \Big)_{(t)} \nonumber \\
& &  \otimes \Big( \vert 0_{1}\ldots 0_{k-1}0_{k} \rangle \Big)_{(t+1)} \bigg] , \nonumber \\
\vert \overline{1} \rangle_{L} & = & \bigg[ \Big( \vert 0_{1} \ldots 0_{k-1}0_{k} \rangle \Big)_{(0)} \otimes \ldots \otimes \Big( \vert 0_{1}\ldots 0_{k-1}0_{k} \rangle \Big)_{(t)} \nonumber \\
& & \otimes \Big( \vert 0_{1} \ldots 0_{k-1}1_{k} \rangle \Big)_{(t+1)} \bigg] , \nonumber \\
\vdots & & \nonumber \\
\vert \overline{2^{k}-2} \rangle_{L} & = & \bigg[ \Big( \vert 0_{1}  \ldots 0_{k-1}0_{k} \rangle \Big)_{(0)} \otimes \ldots \otimes \Big( \vert 0_{1} \ldots 0_{k-1}0_{k} \rangle \Big)_{(t)} \nonumber \\
& &  \otimes \Big( \vert 1_{1}\ldots 1_{k-1}0_{k} \rangle \Big)_{(t+1)} \bigg] , \nonumber \\
\vert \overline{2^{k}-1} \rangle_{L} & = & \bigg[ \Big( \vert 0_{1} \ldots  0_{k-1}0_{k} \rangle \Big)_{(0)} \otimes \ldots \otimes \Big( \vert 0_{1} \ldots 0_{k-1}0_{k} \rangle \Big)_{(t)} \nonumber \\
& &  \otimes \Big( \vert 1_{1}\ldots 1_{k-1}1_{k} \rangle \Big)_{(t+1)} \bigg]  .
\end{eqnarray}

We note in (\ref{apldec_errocase2}) that, after applying the operator (\ref{dec_errocase2}), all blocks of index $(0)$ to $(t)$ were transformed from the GHZ basis into the computational basis. After that, these blocks were identically prepared in the block of index $(t +1)$ and had their $k$ qubits transformed into the $\ket{0}$ state. 

This completes the proof of Lemma \ref{opdecghz}.
\begin{flushright}
$\square$
\end{flushright}

After we apply $U_{dec}$ (Lemma \ref{opdecghz}) to $\Big( \vert \overline{\psi} \rangle_{GHZ} \otimes \vert 0^{\otimes k} \rangle_{(t+1)} \Big)$, it is necessary to apply recovery operators, one operator for each block that had somehow detected erasure, in order to obtain the $\ket{\psi}$ state free of erasures.

Let $B \subset D \ (D = \{ 0, \ldots, t \})$ be the set of indices that identify the blocks where erasure occurred. Upon applying the recovery operator, we must consider two cases for the qubit in which erasure occurred in the block of index $(b) \in B$:

\begin{enumerate}
\item When the position is different from $k$.
\item When the position is equal to $k$.
\end{enumerate}


The next lemma shows how it should be the general form of the recovery operator for the Case $1$.

\begin{lemma}\label{recNk}
Let $B \subset D \ (D = \{ 0, \ldots, t \})$ be the set of indices that identify the blocks where erasure occurred and also consider that $U_{dec}$ has been applied  to $\Big( \vert \overline{\psi} \rangle_{GHZ} \otimes \vert 0^{\otimes k} \rangle_{(t+1)} \Big)$. If the position of the qubit is different from $k$, for the qubit in the position  $a$ where erasure occurred in the block of index $(b) \in B$, then the operator  $U_{rec}^{ a, b}$ that will transform the block of index $(b)$ in the canonical GHZ state is given by

\begin{eqnarray}\label{oprecNk}
U_{rec}^{ a, b} & = & T_{[k -r](t+1),k(t+1), r(b)} Z_{k(t+1), r (b) }  \ T_{[k-r](t+1),k(t+1), r(b)}  \nonumber \\
                     & &   \prod_{i=1  (i \neq a )}^{k-1} C_{i(t+1),i(b)}  \prod_{i=1 (i \neq a )}^{k} C_{[k-r](t+1),i(b)},
\end{eqnarray}\\
where $r=\max_{r \neq k} ( \mathcal{W})$ and $\mathcal{W} = \{ 1, \ldots , k \} \setminus \{ a \}$, with $T$ representing a Toffoli gate operation and $Z$ representing the $\sigma_{Z}$-Pauli controlled operation.
\end{lemma}
\noindent{\bf Proof:}  It will be shown that considering $\vert \overline{\psi} \rangle_{GHZ}$, a state that has $(t +1)$ blocks of $k$-qubits each ($k \geq 3$) in the GHZ basis which has $t = \lfloor k / 2 \rfloor$ erasures in different blocks after passing through the QEC, then the restoring operation, given by (\ref{oprecNk}), is able to get the state that was encoded free of erasures when the position of erasure is different from $k$.


The $\ket{\overline{\psi}}_{GHZ}$ state contains $(t+1)$ blocks and has $t$ erasures in different blocks. Since there are $(t+1)$ blocks, the erasures would have occurred in blocks of indices $(0)$ to $(t)$. However, without  loss of generality, we will consider that erasures occurred in any position $a$ ($a \neq k$) in the blocks of indices $(0)$ to $(t - 1)$, i.e., $B=\{ 0, \ldots , t-1 \}$. So, upon applying $U_{dec}$ to $\Big( \vert \overline{\psi} \rangle_{GHZ} \otimes \vert 0^{\otimes k} \rangle_{(t+1)} \Big)$, the $\vert \overline{\psi} \rangle_{GHZ}$ state is rewritten as follows:

\begin{eqnarray}\label{errorNk_estado}
\vert e_{0} \rangle \otimes  \vert \psi \rangle_{GHZ} \rightarrow \vert \overline{\psi} \rangle_{GHZ} & = & \lambda_{0} \vert \overline{0} \rangle_{L} + \lambda_{1} \vert \overline{1} \rangle_{L} +  \ldots + \lambda_{2^{k}-2} \vert \overline{2^{k}-2} \rangle_{L} + \lambda_{2^{k}-1} \vert \overline{2^{k}-1} \rangle_{L}, \nonumber \\
&  &  
\end{eqnarray}\\
where 

\begin{eqnarray}\label{errorNk}
\vert \overline{0} \rangle_{L} & = & \bigg[ \Big( \vert  \ldots \overline{0}_{a} \ldots 0_{k} \rangle \pm \vert  \ldots \overline{1}_{a} \ldots 1_{k} \rangle \Big)_{(0)} \otimes \ldots \nonumber \\
& & \otimes \Big( \vert  \ldots \overline{0}_{a} \ldots 0_{k} \rangle \pm \vert  \ldots \overline{1}_{a} \ldots 1_{k} \rangle \Big)_{(t-1)} \otimes \Big( \vert \ldots 0_{a} \ldots0_{k} \rangle \Big)_{(t)} \nonumber \\
& & \otimes \Big( \vert  \ldots 0_{a} \ldots 0_{k} \rangle  \Big)_{(t+1)} \bigg] , \nonumber \\
\vert \overline{1} \rangle_{L} & = & \bigg[ \Big( \vert  \ldots \overline{0}_{a} \ldots 0_{k} \rangle \mp \vert  \ldots \overline{1}_{a} \ldots 1_{k} \rangle \Big)_{(0)} \otimes \ldots \nonumber  \\
& & \otimes \Big( \vert  \ldots \overline{0}_{a} \ldots 0_{k} \rangle \mp \vert  \ldots \overline{1}_{a} \ldots 1_{k} \rangle \Big)_{(t-1)} \otimes \Big( \vert  \ldots 0_{a} \ldots 0_{k} \rangle  \Big)_{(t)} \nonumber \\
& & \otimes \Big( \vert \ldots 0_{a} \ldots 1_{k} \rangle \Big)_{(t+1)} \bigg] , \cdots , \nonumber \\
\vert \overline{2^{k}-2} \rangle_{L} & = & \bigg[ \Big( \pm \vert  \ldots \overline{1}_{a} \ldots 0_{k} \rangle + \vert  \ldots \overline{0}_{a} \ldots 1_{k} \rangle \Big)_{(0)} \otimes \ldots \nonumber \\
& & \otimes \Big( \pm \vert  \ldots \overline{1}_{a} \ldots 0_{k} \rangle + \vert  \ldots \overline{0}_{a} \ldots 1_{k} \rangle \Big)_{(t-1)} \otimes \Big( \vert  \ldots 0_{a} \ldots 0_{k} \rangle  \Big)_{(t)} \nonumber \\
& & \otimes \Big( \vert \ldots 1_{a} \ldots 0_{k} \rangle \Big)_{(t+1)} \bigg] , \nonumber \\
\vert \overline{2^{k}-1} \rangle_{L} & = & \bigg[ \Big( \pm \vert  \ldots \overline{1}_{a} \ldots 0_{k} \rangle - \vert  \ldots \overline{0}_{a} \ldots 1_{k} \rangle \Big)_{(0)} \otimes \ldots \nonumber \\
& & \otimes \Big( \pm \vert  \ldots \overline{1}_{a} \ldots 0_{k} \rangle - \vert  \ldots \overline{0}_{a} \ldots 1_{k} \rangle \Big)_{(t-1)} \otimes \Big( \vert  \ldots 0_{a} \ldots 0_{k} \rangle  \Big)_{(t)} \nonumber \\
& & \otimes \Big( \vert \ldots 1_{a} \ldots 1_{k} \rangle \Big)_{(t+1)} \bigg] .
\end{eqnarray}

Considering that the erasure occurred in a qubit of position $a \ (1 \leq a \leq k-1)$ of the blocks of indices $(0)$ to $(t-1)$, then $\mathcal{W}=\{1, \ldots , k\} \setminus \{ a \}$ and $r=max_{r\neq k} (\mathcal{W})$. The recovery operators, one operator for each block of indices $(0)$ to $(t-1)$, are explicitly given as follows:

\begin{eqnarray}\label{oprecNkexp}
U_{rec}^{ a, 0} & = & T_{[k -r](t+1),k(t+1), r(0)}  Z_{k(t+1), r (0) }  \ T_{[k-r](t+1),k(t+1), r(0)}  \nonumber \\
                      & &   \prod_{i=1  (i \neq a )}^{k-1} C_{i(t+1),i(0)}  \prod_{i=1 (i \neq a )}^{k} C_{[k-r](t+1),i(0)}, \nonumber \\
\vdots             &  & \nonumber \\
U_{rec}^{ a, t-1} & = & T_{[k -r](t+1),k(t+1), r(t-1)}  Z_{k(t+1), r (t-1) }  \ T_{[k-r](t+1),k(t+1), r(t-1)}  \nonumber \\
                      & &   \prod_{i=1  (i \neq a )}^{k-1} C_{i(t+1),i(t-1)}  \prod_{i=1 (i \neq a )}^{k} C_{[k-r](t+1),i(t-1)} .
\end{eqnarray}

Given that $U_{dec}$ has been applied  to $\Big( \vert \overline{\psi} \rangle_{GHZ} \otimes \vert 0^{\otimes k} \rangle_{(t+1)} \Big)$, then applying the recovery operators, given by (\ref{oprecNkexp}) in (\ref{errorNk_estado}), we obtain

\begin{eqnarray}\label{aplrecNk}
\vert \overline{0} \rangle_{L} & = & \bigg[ \Big( \vert  \ldots \overline{0}_{a} \ldots 0_{k} \rangle \pm \vert  \ldots \overline{1}_{a} \ldots 1_{k} \rangle \Big)_{(0)} \otimes \ldots \nonumber \\
& & \otimes \Big( \vert  \ldots \overline{0}_{a} \ldots 0_{k} \rangle \pm \vert  \ldots \overline{1}_{a} \ldots 1_{k} \rangle \Big)_{(t-1)} \otimes \Big( \vert \ldots 0_{a} \ldots0_{k} \rangle \Big)_{(t)} \nonumber \\
& & \otimes \Big( \vert  \ldots 0_{a} \ldots 0_{k} \rangle  \Big)_{(t+1)} \bigg] , \nonumber \\
\vert \overline{1} \rangle_{L} & = & \bigg[ \Big( \vert  \ldots \overline{0}_{a} \ldots 0_{k} \rangle \pm \vert  \ldots \overline{1}_{a} \ldots 1_{k} \rangle \Big)_{(0)} \otimes \ldots \nonumber  \\
& & \otimes \Big( \vert  \ldots \overline{0}_{a} \ldots 0_{k} \rangle \pm \vert  \ldots \overline{1}_{a} \ldots 1_{k} \rangle \Big)_{(t-1)} \otimes \Big( \vert  \ldots 0_{a} \ldots 0_{k} \rangle  \Big)_{(t)} \nonumber \\
& & \otimes \Big( \vert \ldots 0_{a} \ldots 1_{k} \rangle \Big)_{(t+1)} \bigg] , \cdots , \nonumber \\
\vert \overline{2^{k}-2} \rangle_{L} & = & \bigg[ \Big( \vert  \ldots \overline{0}_{a} \ldots 0_{k} \rangle \pm \vert  \ldots \overline{1}_{a} \ldots 1_{k} \rangle \Big)_{(0)} \otimes \ldots \nonumber  \\
& & \otimes \Big( \vert  \ldots \overline{0}_{a} \ldots 0_{k} \rangle \pm \vert  \ldots \overline{1}_{a} \ldots 1_{k} \rangle \Big)_{(t-1)} \otimes \Big( \vert  \ldots 0_{a} \ldots 0_{k} \rangle  \Big)_{(t)} \nonumber \\
& & \otimes \Big( \vert \ldots 1_{a} \ldots 0_{k} \rangle \Big)_{(t+1)} \bigg] , \nonumber \\
\vert \overline{2^{k}-1} \rangle_{L} & = & \bigg[ \Big( \vert  \ldots \overline{0}_{a} \ldots 0_{k} \rangle \pm \vert  \ldots \overline{1}_{a} \ldots 1_{k} \rangle \Big)_{(0)} \otimes \ldots \nonumber  \\
& & \otimes \Big( \vert  \ldots \overline{0}_{a} \ldots 0_{k} \rangle \pm \vert  \ldots \overline{1}_{a} \ldots 1_{k} \rangle \Big)_{(t-1)} \otimes \Big( \vert  \ldots 0_{a} \ldots 0_{k} \rangle  \Big)_{(t)} \nonumber \\
& & \otimes \Big( \vert \ldots 1_{a} \ldots 1_{k} \rangle \Big)_{(t+1)} \bigg] . 
\end{eqnarray}\\
Notice now in (\ref{aplrecNk}) that the blocks of indices $(0)$ to $(t-1)$ are in a canonical GHZ state. This way, the system and the environment will be in the state

\begin{eqnarray}\label{recsit2i}
& & \Big( \vert  \ldots \overline{x}_{a} \ldots 0_{k} \rangle \pm \vert  \ldots \overline{x}_{a} \ldots 1_{k} \rangle \Big)_{(0)} \otimes \Big( \vert  \ldots \overline{x}_{a} \ldots 0_{k} \rangle \pm \vert  \ldots \overline{x}_{a} \ldots 1_{k} \rangle \Big)_{(1)}   \nonumber \\
& & \otimes \ldots \otimes \Big( \vert  \ldots \overline{x}_{a} \ldots 0_{k} \rangle \pm \vert  \ldots \overline{x}_{a} \ldots 1_{k} \rangle \Big)_{(t-1)} \otimes \Big( \vert \ldots 0_{a} \ldots0_{k} \rangle \Big)_{(t)}  \nonumber \\
& & \otimes \Big( \vert  \psi \rangle  \Big)_{(t+1)} ,
\end{eqnarray}\\
where $\overline{x}_{a} \in \{ 0 ,1 \}$.  

Thus, the original message state $\ket{\psi}$ can be recovered via the block of index $(t+1)$ even after passing through the QEC and also after the occurrence of erasure in the qubit of position $\{ a \} \ (a \neq k)$  of the blocks of indices $(0)$ to $(t-1)$.
 
\begin{flushright}
$\square$
\end{flushright}


After the proof of the Lemma \ref{recNk} for the Case 1, now the Case 2 and its associated lemma are considered.

\begin{lemma}\label{rec_posik}
Consider $B \subset D \ (D = \{ 0, \ldots, t \})$ as the set of indices that identify the blocks where erasures occurred and to which the $U_{dec}$ operator has been applied to $\Big( \vert \overline{\psi} \rangle_{GHZ} \otimes \vert 0^{\otimes k} \rangle_{(t+1)} \Big)$. If the position is equal to $k$, for the qubit in which erasure occurred in the block of index $(b) \in B$, then the $U_{rec}^{ k, b}$ operator that will transform the  index block $(b)$ into a canonical GHZ state is given by

\begin{eqnarray}\label{oprecposik}
U_{rec}^{ k, b} = Z_{k(t+1), k-1 (b) }   \prod_{i=1}^{k-1} C_{i(t+1), i(b)},   
\end{eqnarray}\\
where $Z$ representing the $\sigma_{Z}$-Pauli controlled operation.
\end{lemma}

\noindent{\bf Proof:} Let the considerations to  $\vert \overline{\psi} \rangle_{GHZ}$, $t$, $k$ and to the number of blocks be the same of the Lemma \ref{recNk}, except that the position of erasure is equal to $k$. So, the $\vert \overline{\psi} \rangle_{GHZ}$ state for this situation, considering that $U_{dec}$ has been applied  to $\Big( \vert \overline{\psi} \rangle_{GHZ} \otimes \vert 0^{\otimes k} \rangle_{(t+1)} \Big)$, has the following form:

\begin{eqnarray}\label{errork_estado}
\vert e_{0} \rangle \otimes  \vert \psi \rangle_{GHZ} \rightarrow \vert \overline{\psi} \rangle_{GHZ} & = & \lambda_{0} \vert \overline{0} \rangle_{L} + \lambda_{1} \vert \overline{1} \rangle_{L} +  \ldots + \lambda_{2^{k}-2} \vert \overline{2^{k}-2} \rangle_{L} + \lambda_{2^{k}-1} \vert \overline{2^{k}-1} \rangle_{L}, \nonumber \\
&  &  
\end{eqnarray}\\
where 

\begin{eqnarray}\label{error_k}
\vert \overline{0} \rangle_{L} & = & \bigg[ \Big( \vert 0 \ldots 0_{k-1} \overline{0}_{k} \rangle \pm \vert 1 \ldots 1_{k-1} \overline{1}_{k} \rangle \Big)_{(0)} \otimes \ldots \nonumber \\
& & \otimes \Big( \vert 0 \ldots 0_{k-1} \overline{0}_{k} \rangle \pm \vert 1 \ldots 1_{k-1} \overline{1}_{k} \rangle \Big)_{(t-1)} \otimes \Big( \vert 0 \ldots 0_{k-1} 0_{k} \rangle \Big)_{(t)} \nonumber \\
& & \otimes \Big( \vert 0 \ldots 0_{k-1} 0_{k} \rangle  \Big)_{(t+1)} \bigg] , \nonumber \\
\vert \overline{1} \rangle_{L} & = & \bigg[ \Big( \vert 0 \ldots 0_{k-1} \overline{0}_{k} \rangle \mp \vert 1 \ldots 1_{k-1} \overline{1}_{k} \rangle \Big)_{(0)} \otimes \ldots \nonumber \\
& & \otimes \Big( \vert 0 \ldots 0_{k-1} \overline{0}_{k} \rangle \mp \vert 1 \ldots 1_{k-1} \overline{1}_{k} \rangle \Big)_{(t-1)} \otimes \Big( \vert 0 \ldots 0_{k-1} 0_{k} \rangle \Big)_{(t)} \nonumber \\
& & \otimes \Big( \vert 0 \ldots 0_{k-1} 1_{k} \rangle  \Big)_{(t+1)} \bigg] , \cdots , \nonumber \\
\vert \overline{2^{k}-2} \rangle_{L} & = & \bigg[ \Big( \pm \vert 1  \ldots 1_{k-1} \overline{0}_{k} \rangle + \vert 0 \ldots 0_{k-1} \overline{1}_{k} \rangle \Big)_{(0)} \otimes \ldots \nonumber \\
& & \otimes  \Big( \pm \vert 1  \ldots 1_{k-1} \overline{0}_{k} \rangle + \vert 0 \ldots 0_{k-1} \overline{1}_{k} \rangle \Big)_{(t-1)} \otimes \Big( \vert 0 \ldots 0_{k-1} 0_{k} \rangle  \Big)_{(t)} \nonumber \\
& & \otimes \Big( \vert 1 \ldots 1_{k-1} 0_{k} \rangle \Big)_{(t+1)} \bigg] , \nonumber \\
\vert \overline{2^{k}-1} \rangle_{L} & = & \bigg[ \Big( \pm \vert 1 \ldots 1_{k-1} \overline{0}_{k} \rangle - \vert 0 \ldots 0_{k-1} \overline{1}_{k} \rangle \Big)_{(0)} \otimes \ldots \nonumber \\
& & \otimes \Big( \pm \vert 1 \ldots 1_{k-1} \overline{0}_{k} \rangle - \vert 0 \ldots 0_{k-1} \overline{1}_{k} \rangle \Big)_{(t-1)} \otimes \Big( \vert 0 \ldots 0_{k-1} 0_{k} \rangle  \Big)_{(t)} \nonumber \\
& & \otimes \Big( \vert 1 \ldots 1_{k-1} 1_{k} \rangle \Big)_{(t+1)} \bigg] .
\end{eqnarray}

Considering that the erasure occurred in the qubit of position $k$ of the blocks of indices $(0)$ to $(t-1)$, then $\mathcal{W}=\{1, \ldots , k\} \setminus \{ k \} = \{ 1, \ldots , k -1 \}$ and $r=max_{r\neq k} (\mathcal{W})=k-1$. The recovery operators, one operator for each block of indices $(0)$ to $(t-1)$, are explicitly given as follows:

\begin{eqnarray}\label{opreckexp}
U_{rec}^{ k, 0} & = & Z_{k(t+1), k-1 (0) }  \prod_{i=1}^{k-1} C_{i(t+1),i(0)} , \nonumber \\
\vdots             &  & \nonumber \\
U_{rec}^{ k, t-1} & = & Z_{k(t+1), k-1 (t-1) }  \prod_{i=1}^{k-1} C_{i(t+1),i(t-1)}  .
\end{eqnarray}

Applying the recovery operators given by (\ref{opreckexp}) in (\ref{errork_estado}), we obtain

\begin{eqnarray}\label{aplrec_k}
\vert \overline{0} \rangle_{L} & = & \bigg[ \Big( \vert 0 \ldots 0_{k-1} \overline{0}_{k} \rangle \pm \vert 1 \ldots 1_{k-1} \overline{1}_{k} \rangle \Big)_{(0)} \otimes \ldots \nonumber \\
& & \otimes \Big( \vert 0 \ldots 0_{k-1} \overline{0}_{k} \rangle \pm \vert 1 \ldots 1_{k-1} \overline{1}_{k} \rangle \Big)_{(t-1)} \otimes \Big( \vert 0 \ldots 0_{k-1} 0_{k} \rangle \Big)_{(t)} \nonumber \\
& & \otimes \Big( \vert 0 \ldots 0_{k-1} 0_{k} \rangle  \Big)_{(t+1)} \bigg] , \nonumber \\
\vert \overline{1} \rangle_{L} & = & \bigg[ \Big( \vert 0 \ldots 0_{k-1} \overline{0}_{k} \rangle \pm \vert 1 \ldots 1_{k-1} \overline{1}_{k} \rangle \Big)_{(0)} \otimes \ldots \nonumber \\
& & \otimes \Big( \vert 0 \ldots 0_{k-1} \overline{0}_{k} \rangle \pm \vert 1 \ldots 1_{k-1} \overline{1}_{k} \rangle \Big)_{(t-1)} \otimes \Big( \vert 0 \ldots 0_{k-1} 0_{k} \rangle \Big)_{(t)} \nonumber \\
& & \otimes \Big( \vert 0 \ldots 0_{k-1} 1_{k} \rangle  \Big)_{(t+1)} \bigg] , \cdots , \nonumber \\
\vert \overline{2^{k}-2} \rangle_{L} & = & \bigg[ \Big( \vert 0 \ldots 0_{k-1} \overline{0}_{k} \rangle \pm \vert 1 \ldots 1_{k-1} \overline{1}_{k} \rangle \Big)_{(0)} \otimes \ldots \nonumber \\
& & \otimes  \Big( \vert 0 \ldots 0_{k-1} \overline{0}_{k} \rangle \pm \vert 1 \ldots 1_{k-1} \overline{1}_{k} \rangle \Big)_{(t-1)} \otimes \Big( \vert 0 \ldots 0_{k-1} 0_{k} \rangle  \Big)_{(t)} \nonumber \\
& & \otimes \Big( \vert 1 \ldots 1_{k-1} 0_{k} \rangle \Big)_{(t+1)} \bigg] , \nonumber \\
\vert \overline{2^{k}-1} \rangle_{L} & = & \bigg[ \Big( \vert 0 \ldots 0_{k-1} \overline{0}_{k} \rangle \pm \vert 1 \ldots 1_{k-1} \overline{1}_{k} \rangle \Big)_{(0)} \otimes \ldots \nonumber \\
& & \otimes \Big( \vert 0 \ldots 0_{k-1} \overline{0}_{k} \rangle \pm \vert 1 \ldots 1_{k-1} \overline{1}_{k} \rangle \Big)_{(t-1)} \otimes \Big( \vert 0 \ldots 0_{k-1} 0_{k} \rangle  \Big)_{(t)} \nonumber \\
& & \otimes \Big( \vert 1 \ldots 1_{k-1} 1_{k} \rangle \Big)_{(t+1)} \bigg] .
\end{eqnarray}\\
Notice now in (\ref{aplrec_k}) that the blocks of indices $(0)$ to $(t-1)$ are in the canonical GHZ state. This way, the system and the environment will be in the state

\begin{eqnarray}\label{recsit2i}
& & \Big( \vert 0 \ldots 0_{k-1} \overline{0}_{k} \rangle \pm \vert 1 \ldots 1_{k-1} \overline{1}_{k} \rangle \Big)_{(0)} \otimes \Big( \vert 0 \ldots 0_{k-1} \overline{0}_{k} \rangle \pm \vert 1 \ldots 1_{k-1} \overline{1}_{k} \rangle \Big)_{(1)}   \nonumber \\
& & \otimes \ldots \otimes \Big( \vert 0 \ldots 0_{k-1} \overline{0}_{k} \rangle \pm \vert 1 \ldots 1_{k-1} \overline{1}_{k} \rangle \Big)_{(t-1)} \otimes \Big( \vert 0 \ldots 0_{k-1} 0_{k} \rangle \Big)_{(t)}  \nonumber \\
& & \otimes \Big( \vert  \psi \rangle  \Big)_{(t+1)} .
\end{eqnarray}  

Thus, the original message state $\ket{\psi}$ can be recovered via the block of index $(t+1)$, even after passing through the QEC and also  after occurrence of erasure in qubit of position $k$ of the blocks of indices $(0)$ to $(t-1)$.

\begin{flushright}
$\square$
\end{flushright}


Consider that the state $\vert \psi \rangle_{GHZ}$ passed through the QEC and that $t = \lfloor k / 2 \rfloor$ erasures occurred, resulting in $\vert \overline{\psi} \rangle_{GHZ}$. The following theorem shows how the restoring operation most be performed in order to recover the encoded state free of erasures.

\begin{theorem}\label{restoper}
Let $\vert \overline{\psi} \rangle_{GHZ}$ be a state that has $(t+1)$ redundant blocks of $k$-qubits ($k \geq 3$) each in the GHZ basis in which $t = \lfloor k / 2 \rfloor$ erasures occurred after passing through the QEC, and let $B \subset D \ (D = \{ 0, \ldots, t \})$ be the set of indices that identify the blocks in which there were detected erasures, then the restoring operation $\mathcal{R}$ able to recover the original state, is given by
  
\begin{eqnarray}\label{oprecov2}
\mathcal{R} & = & \prod_{b} \Bigg\{ U_{rec}^{ a, b} \circ \Big[ U_{dec} \Big( \vert \overline{\psi} \rangle_{GHZ} \otimes \vert 0^{\otimes k} \rangle_{(t+1)} \Big) \Big] \Bigg\} ,
\end{eqnarray} \\
where $a \in \{1, \ldots , k \}$ is the position of the qubit that suffered erasure and $b \in B$, with $U_{dec}$ as in (\ref{opdecod1}) and $U_{rec}^{ a, b}$ be as in (\ref{oprecNk}) or as in (\ref{oprecposik}), except for $U_{rec}^{0}=I$ when $|B|=0$ (where $I$ is the identity matrix of order $k$).
 \end{theorem}

\noindent{\bf Proof:} It will be shown that taking $\vert \overline{\psi} \rangle_{GHZ}$, a state that has $(t +1)$ redundant blocks of $k$-qubits each ($k \geq 3$) in the GHZ basis, which has $t = \lfloor k / 2 \rfloor$ erasures after passing through the QEC, then the restoring operation, given by (\ref{oprecov2}), is able to recover the originally encoded state free of erasures.

The demonstration must consider both cases involving the decoding operator (see Lemma \ref{opdecghz}), and also the cases involving the recovery operator (see Lemmas \ref{recNk} and \ref{rec_posik}), that is:

\begin{itemize}
\item all blocks remain undamaged (no erasure); 
\item just one block stay undamaged ($t$ erasures occurred); 
\item the position is different from the last ($k$-th position) for the qubit who suffered erasure;
\item the position is equal to $k$ for the qubit that suffered erasure.
\end{itemize}

For the first case listed above, the proof is given by Lemma  \ref{opdecghz}. To demonstrate the other three cases the following situation will be presented.

We will consider that the $\vert \overline{\psi} \rangle_{GHZ}$ state contains ($t +1$) blocks and suffered $t = \lfloor k / 2 \rfloor$ erasures in different blocks after passing through the QEC. Since there are ($t +1$) blocks, so there will be a block in which no erasure was detected (one block will stay undamaged). This block can be any of the ($t +1$) blocks, including blocks of indices $(0)$ and $(t)$. However, without loss of generality, we will assume that the undamaged block is the index block ($t$). Because of this, erasures will occur in any position $\{ a \}$ in the blocks of indices $(0)$ to $(t-1)$.

In order to deal with applications of the recovery operator given by Lemmas \ref{recNk} and \ref{rec_posik}, we will consider that the position of the qubit that suffered erasure is the $k$-th in the index block $(0)$ and different of the $k$-th in the blocks of indices $(1)$ to $(t-1)$. The $\vert \overline{\psi} \rangle_{GHZ}$ state for this situation has the following form:

\begin{eqnarray}\label{sit1estado}
\vert e_{0} \rangle \otimes  \vert \psi \rangle_{GHZ} \rightarrow \vert \overline{\psi} \rangle_{GHZ} & = & \lambda_{0} \vert \overline{0} \rangle_{L} + \lambda_{1} \vert \overline{1} \rangle_{L} +  \ldots + \lambda_{2^{k}-2} \vert \overline{2^{k}-2} \rangle_{L} + \lambda_{2^{k}-1} \vert \overline{2^{k}-1} \rangle_{L}, \nonumber \\
&  & 
\end{eqnarray}
where

\begin{eqnarray}\label{sit1error}
\vert \overline{0} \rangle_{L} & = & \bigg[ \Big( \vert  \ldots 0_{a} \ldots \overline{0}_{k} \rangle \pm \vert  \ldots 1_{a} \ldots \overline{1}_{k} \rangle \Big)_{(0)} \otimes \ldots \nonumber \\
& & \otimes \Big( \vert  \ldots \overline{0}_{a} \ldots 0_{k} \rangle \pm \vert  \ldots \overline{1}_{a} \ldots 1_{k} \rangle \Big)_{(t-1)} \otimes \nonumber \\
& & \otimes \Big( \vert \ldots 0_{a} \ldots0_{k} \rangle + \vert \ldots 1_{a} \ldots 1_{k} \rangle \Big)_{(t)} \bigg] , \nonumber \\
\vert \overline{1} \rangle_{L} & = & \bigg[ \Big( \vert  \ldots 0_{a} \ldots \overline{0}_{k} \rangle \mp \vert  \ldots 1_{a} \ldots \overline{1}_{k} \rangle \Big)_{(0)} \otimes \ldots \nonumber \\
& & \otimes \Big( \vert  \ldots \overline{0}_{a} \ldots 0_{k} \rangle \mp \vert  \ldots \overline{1}_{a} \ldots 1_{k} \rangle \Big)_{(t-1)} \nonumber \\
& & \otimes \Big( \vert \ldots 0_{a} \ldots 0_{k} \rangle - \vert  \ldots 1_{a} \ldots 1_{k} \rangle \Big)_{(t)} \bigg] , \cdots , \ \nonumber \\
\vert \overline{2^{k}-2} \rangle_{L} & = & \bigg[ \Big( \vert  \ldots 1_{a} \ldots \overline{0}_{k} \rangle \pm \vert  \ldots 0_{a} \ldots \overline{1}_{k} \rangle \Big)_{(0)} \otimes \ldots \nonumber \\
& & \otimes \Big( \pm \vert  \ldots \overline{1}_{a} \ldots  0_{k} \rangle + \vert  \ldots \overline{0}_{a} \ldots 1_{k} \rangle \Big)_{(t-1)} \nonumber \\
& & \otimes \Big( \vert \ldots 1_{a} \ldots 0_{k} \rangle + \vert  \ldots 0_{a} \ldots 1_{k} \rangle \Big)_{(t)} \bigg] , \nonumber \\
\vert \overline{2^{k}-1} \rangle_{L} & = & \bigg[ \Big(  \vert  \ldots 1_{a} \ldots \overline{0}_{k} \rangle \mp \vert  \ldots 0_{a} \ldots \overline{1}_{k} \rangle \Big)_{(0)} \otimes \ldots  \nonumber \\
& & \otimes \Big( \pm \vert  \ldots \overline{1}_{a} \ldots 0_{k} \rangle - \vert  \ldots \overline{0}_{a} \ldots  1_{k} \rangle \Big)_{(t-1)} \nonumber \\
& & \otimes \Big( \vert \ldots 1_{a} \ldots 0_{k} \rangle - \vert  \ldots 0_{a} \ldots 1_{k} \rangle \Big)_{(t)} \bigg] . 
\end{eqnarray}

The restoring operation is, therefore, given as follows:

\begin{eqnarray}\label{oprestsit1}
 \mathcal{R} & = & \Bigg\{ U_{rec}^{k, 0} \circ U_{dec} \Big( \vert \overline{\psi} \rangle_{GHZ} \otimes \vert 0^{\otimes k} \rangle_{(t+1)} \Big) \Bigg\} \nonumber \\
 & & \Bigg\{  \prod_{b=1}^{t-1} \bigg[ U_{rec}^{  a, b} \circ U_{dec} \Big( \vert \overline{\psi} \rangle_{GHZ} \otimes \vert 0^{\otimes k} \rangle_{(t+1)} \Big) \bigg] \Bigg\} .
\end{eqnarray}

The first step in the operation given in (\ref{oprestsit1}) is the application of $U_{dec}$ which will only act in the block of index $(t)$.

Applying $U_{dec}$ to the product $\Big( \vert \overline{\psi} \rangle_{GHZ} \otimes \vert 0^{\otimes k} \rangle_{(t+1)} \Big)$, we obtain

\begin{eqnarray}\label{sit1decerror}
\vert \overline{0} \rangle_{L} & = & \bigg[ \Big( \vert  \ldots 0_{a} \ldots \overline{0}_{k} \rangle \pm \vert  \ldots 1_{a} \ldots \overline{1}_{k} \rangle \Big)_{(0)} \otimes \ldots \nonumber \\
& & \otimes \Big( \vert  \ldots \overline{0}_{a} \ldots 0_{k} \rangle \pm \vert  \ldots \overline{1}_{a} \ldots 1_{k} \rangle \Big)_{(t-1)} \otimes \Big( \vert \ldots 0_{a} \ldots 0_{k} \rangle \Big)_{(t)} \nonumber \\
& &  \otimes \Big( \vert  \ldots 0_{a} \ldots 0_{k} \rangle  \Big)_{(t+1)} \bigg] , \nonumber \\
\vert \overline{1} \rangle_{L} & = & \bigg[ \Big( \vert  \ldots 0_{a} \ldots \overline{0}_{k} \rangle \mp \vert  \ldots 1_{a} \ldots \overline{1}_{k} \rangle \Big)_{(0)} \otimes \ldots \nonumber \\
& & \otimes \Big( \vert  \ldots \overline{0}_{a} \ldots 0_{k} \rangle \mp \vert  \ldots \overline{1}_{a} \ldots 1_{k} \rangle \Big)_{(t-1)} \otimes \Big( \vert  \ldots 0_{a} \ldots 0_{k} \rangle  \Big)_{(t)} \nonumber \\
& & \otimes \Big( \vert \ldots 0_{a} \ldots 1_{k} \rangle \Big)_{(t+1)} \bigg] , \cdots , \nonumber \\
\vert \overline{2^{k}-2} \rangle_{L} & = & \bigg[ \Big( \vert  \ldots 1_{a} \ldots \overline{0}_{k} \rangle \pm \vert  \ldots 0_{a} \ldots \overline{1}_{k} \rangle \Big)_{(0)} \otimes \ldots \nonumber \\
& & \otimes \Big( \pm \vert  \ldots \overline{1}_{a} \ldots  0_{k} \rangle + \vert  \ldots \overline{0}_{a} \ldots 1_{k} \rangle \Big)_{(t-1)} \otimes \Big( \vert  \ldots 0_{a} \ldots 0_{k} \rangle  \Big)_{(t)} \nonumber \\
& & \otimes \Big( \vert \ldots 1_{a} \ldots 0_{k} \rangle \Big)_{(t+1)} \bigg] , \nonumber \\
\vert \overline{2^{k}-1} \rangle_{L} & = & \bigg[ \Big( \vert  \ldots 1_{a} \ldots \overline{0}_{k} \rangle \mp \vert  \ldots 0_{a} \ldots \overline{1}_{k} \rangle \Big)_{(0)} \otimes \ldots \nonumber \\ 
& & \otimes \Big( \pm \vert  \ldots \overline{1}_{a} \ldots 0_{k} \rangle - \vert  \ldots \overline{0}_{a} \ldots  1_{k} \rangle \Big)_{(t-1)} \otimes \Big( \vert  \ldots 0_{a} \ldots 0_{k} \rangle  \Big)_{(t)} \nonumber \\
& & \otimes \Big( \vert \ldots 1_{a} \ldots 1_{k} \rangle \Big)_{(t+1)} \bigg] . 
\end{eqnarray}\\
Notice that after the application of $U_{dec}$: (a) the block of index $(t)$ was transformed from the GHZ basis into the computational basis; (b) it was identically prepared in the block of index ($t +1$); and (c) it had all qubits transformed into the $\ket{0}$ state. Also note that there was no change in the blocks of indices $(0)$ to $(t-1)$.

The next step in the operation given in (\ref {oprestsit1}) is the application of recovery operators, one for each block in which erasures were detected. 

For the block of index $(0)$, where the qubit of position $k$ was erased, the recovery operator is given by Lemma \ref{rec_posik}. Considering that, in this case, $\mathcal{W}=\{1, \ldots , k\} \setminus \{ k \}$ and hence $r=k-1$, then it has the following form:

\begin{eqnarray}\label{oprecsit1k}
U_{rec}^{ k, 0} & = & Z_{k(t+1), [k-1] (0) }   \prod_{i=1}^{k-1} C_{i(t+1),i(0)}; \nonumber \\
                       & = & Z_{k(t+1), [k-1] (0) }  \Big(  C_{1(t+1), 1(0)} \cdots C_{[k-1](t+1),[k-1](0)} \Big) .
\end{eqnarray}

For the case where the erasure occurred in a qubit of position $a$ ($1 \leq a \leq k-1$) of the blocks of indices $(1)$ to $(t-1)$, the recovery operator is given by Lemma \ref{recNk}. For this case $\mathcal{W}=\{1, \ldots , k\} \setminus \{ a \}$ and $r=\max_{r \neq k} ( \mathcal{W})$. 
The recovery operators, one operator for each block of indices $(1)$ to $(t-1)$, are explicitly given as follows:

\begin{eqnarray}\label{oprecsit1Nk}
U_{rec}^{ a, 1} & = & T_{[k -r](t+1),k(t+1), r(1)}  Z_{k(t+1), r (1) }  \ T_{[k-r](t+1),k(t+1), r(1)}  \nonumber \\
                      & &   \prod_{i=1  (i \neq a )}^{k-1} C_{i(t+1),i(1)}  \prod_{i=1 (i \neq a )}^{k} C_{[k-r](t+1),i(1)}, \nonumber \\
\vdots             &  & \nonumber \\
U_{rec}^{ a, t-1} & = & T_{[k -r](t+1),k(t+1), r(t-1)}  Z_{k(t+1), r (t-1) }  \ T_{[k-r](t+1),k(t+1), r(t-1)}  \nonumber \\
                      & &   \prod_{i=1  (i \neq a )}^{k-1} C_{i(t+1),i(t-1)}  \prod_{i=1 (i \neq a )}^{k} C_{[k-r](t+1),i(t-1)} .
\end{eqnarray}

Applying the recovery operators, given by (\ref{oprecsit1k}) and (\ref{oprecsit1Nk}) in (\ref{sit1decerror}), we obtain

\begin{eqnarray}\label{sit1aplrec}
\vert \overline{0} \rangle_{L} & = & \bigg[ \Big( \vert  \ldots 0_{a} \ldots \overline{0}_{k} \rangle \pm \vert  \ldots 1_{a} \ldots \overline{1}_{k} \rangle \Big)_{(0)} \otimes \ldots \nonumber \\
& & \otimes \Big( \vert  \ldots \overline{0}_{a} \ldots 0_{k} \rangle \pm \vert  \ldots \overline{1}_{a} \ldots 1_{k} \rangle \Big)_{(t-1)} \otimes \Big( \vert \ldots 0_{a} \ldots0_{k} \rangle \Big)_{(t)} \nonumber \\
& & \otimes \Big( \vert  \ldots 0_{a} \ldots 0_{k} \rangle  \Big)_{(t+1)} \bigg] , \nonumber \\
\vert \overline{1} \rangle_{L} & = & \bigg[ \Big( \vert  \ldots 0_{a} \ldots \overline{0}_{k} \rangle \pm \vert  \ldots 1_{a} \ldots \overline{1}_{k} \rangle \Big)_{(0)} \otimes \ldots \nonumber  \\
& & \otimes \Big( \vert  \ldots \overline{0}_{a} \ldots 0_{k} \rangle \pm \vert  \ldots \overline{1}_{a} \ldots 1_{k} \rangle \Big)_{(t-1)} \otimes \Big( \vert  \ldots 0_{a} \ldots 0_{k} \rangle  \Big)_{(t)} \nonumber \\
& & \otimes \Big( \vert \ldots 0_{a} \ldots 1_{k} \rangle \Big)_{(t+1)} \bigg] , \cdots , \nonumber \\
\vert \overline{2^{k}-2} \rangle_{L} & = & \bigg[ \Big( \vert  \ldots 0_{a} \ldots \overline{0}_{k} \rangle \pm \vert  \ldots 1_{a} \ldots \overline{1}_{k} \rangle \Big)_{(0)} \otimes \ldots \nonumber \\
& & \otimes \Big( \vert  \ldots \overline{0}_{a} \ldots 0_{k} \rangle \pm \vert  \ldots \overline{1}_{a} \ldots 1_{k} \rangle \Big)_{(t-1)} \otimes \Big( \vert  \ldots 0_{a} \ldots 0_{k} \rangle  \Big)_{(t)} \nonumber \\
& & \otimes \Big( \vert \ldots 1_{a} \ldots 0_{k} \rangle \Big)_{(t+1)} \bigg] , \nonumber \\
\vert \overline{2^{k}-1} \rangle_{L} & = & \bigg[ \Big( \vert  \ldots 0_{a} \ldots \overline{0}_{k} \rangle \pm \vert  \ldots 1_{a} \ldots \overline{1}_{k} \rangle \Big)_{(0)} \otimes \ldots \nonumber \\
& & \otimes \Big( \vert  \ldots \overline{0}_{a} \ldots 0_{k} \rangle \pm \vert  \ldots \overline{1}_{a} \ldots 1_{k} \rangle \Big)_{(t-1)} \otimes \Big( \vert  \ldots 0_{a} \ldots 0_{k} \rangle  \Big)_{(t)} \nonumber \\
& & \otimes \Big( \vert \ldots 1_{a} \ldots 1_{k} \rangle \Big)_{(t+1)} \bigg] . 
\end{eqnarray}\\
Notice now in (\ref{sit1aplrec}) that the blocks of indices $(0)$ to $(t-1)$ are in the canonical GHZ state. This way, the system and the environment will be in the state

\begin{eqnarray}\label{recsit2i}
& & \Big( \vert  \ldots 0_{a} \ldots \overline{x}_{k} \rangle + \vert  \ldots 1_{a} \ldots \overline{x}_{k} \rangle \Big)_{(0)} \otimes \Big( \vert  \ldots \overline{x}_{a} \ldots 0_{k} \rangle + \vert  \ldots \overline{x}_{a} \ldots 1_{k} \rangle \Big)_{(1)}   \nonumber \\
& & \otimes \ldots \otimes \Big( \vert  \ldots \overline{x}_{a} \ldots 0_{k} \rangle + \vert  \ldots \overline{x}_{a} \ldots 1_{k} \rangle \Big)_{(t-1)} \otimes \Big( \vert \ldots 0_{a} \ldots0_{k} \rangle \Big)_{(t)}  \nonumber \\
& & \otimes \Big( \vert  \psi \rangle  \Big)_{(t+1)} ,
\end{eqnarray}\\
where $\overline{x}_{a} \in \{ 0 ,1 \}$.  

Thus, the original message state $\ket{\psi}$ can be recovered via the block of index $(t+1)$, even after passing through the QEC, and also after the occurrence of erasure in blocks of indices $(0)$ to $(t-1)$.
 
We conclude, thus, the proof of Theorem \ref{restoper}.
\begin{flushright}
$\square$
\end{flushright}

Aiming at providing a better understanding of the proposed scheme against the occurrence of multiple erasures, an example that illustrates the application of Theorems \ref{encapag} and \ref{restoper} will be presented.

\section{Example}\label{examplescheme}

This example illustrates how the proposed scheme protects 5-qubit of information against $t= \lfloor 5 / 2 \rfloor = 2$ erasures. Along this example, the normalizations factors will be omitted.

For an arbitrary state of $5$ qubits, its binary decomposition is written as follows:

\begin{eqnarray}\label{5qbits}
\vert \psi \rangle & = & \lambda_{0} \vert 00000 \rangle + \lambda_{1} \vert 00001 \rangle + \lambda_{2} \vert 00010 \rangle  + \lambda_{3} \vert 00011 \rangle + \cdots  \nonumber \\
            & & + \lambda_{28} \vert 11100 \rangle + \lambda_{29} \vert 11101 \rangle  + \lambda_{30} \vert 11110 \rangle + \lambda_{31} \vert 11111 \rangle , 
\end{eqnarray}\\
where $\sum_{i=0}^{31} |\lambda_{i}|^{2}=1$.


According to our scheme, the  $\vert \psi \rangle$ state must be encoded in $t+1=3$ blocks of $k=5$ qubits each. The encoding is given as follows:

\begin{equation}\label{enc5}
\mathfrak{E}_{GHZ} = U_{enc}\big( \vert \psi \rangle_{(0)} \otimes \vert 00000 \rangle_{(1)}  \otimes \vert 00000 \rangle_{(2)}\big) ,
\end{equation}
where

\begin{eqnarray}\label{openc5}
U_{enc} & = & \prod_{d=0}^{2} \left( \prod_{i=1}^{4} C_{5(d),i(d)} \right)  \prod_{d=0}^{2} \Bigg( H_{5(d)} \Bigg)  \prod_{d=1}^{2} \left( \prod_{i=1}^{5} C_{i(0),i(d)} \right) \nonumber \\
 & = & \Big( C_{5(0),1(0)}C_{5(0),2(0)}C_{5(0),3(0)}C_{5(0),4(0)} \Big) \nonumber \\
 & & \Big(  C_{5(1),1(1)}C_{5(1),2(1)}C_{5(1),3(1)}C_{5(1),4(1)} \Big)  \nonumber \\
 & & \Big( C_{5(2),1(2)}C_{5(2),2(2)}C_{5(2),3(2)} C_{5(2),4(2)} \Big)  \nonumber \\
 & & \Big(   H_{5(0)}H_{5(1)}H_{5(2)} \Big) \nonumber \\
 & & \Big( C_{1(0),1(1)} C_{2(0),2(1)}C_{3(0),3(1)}C_{4(0),4(1)}C_{5(0),5(1)}   \Big) \nonumber \\
 & & \Big( C_{1(0),1(2)} C_{2(0),2(2)}C_{3(0),3(2)}  C_{4(0),4(2)}C_{5(0),5(2)} \Big) . 
\end{eqnarray}

Applying the operator (\ref{openc5}) in the product  $\big( \vert \psi \rangle_{(0)} \otimes \vert 00000 \rangle_{(1)}  \otimes \vert 00000 \rangle_{(2)}\big)$, we obtain

\begin{eqnarray}\label{fiveancil}
\vert \psi \rangle_{GHZ} & = & \lambda_{0} \vert 0 \rangle_{L} + \lambda_{1} \vert 1 \rangle_{L} + \lambda_{2} \vert 2 \rangle_{L} + \lambda_{3} \vert 3 \rangle_{L} + \cdots + \lambda_{28} \vert 28 \rangle_{L}   \nonumber \\
 & & + \lambda_{29} \vert 29 \rangle_{L} + \lambda_{30} \vert 30 \rangle_{L} + \lambda_{31}  \vert 31 \rangle_{L},
\end{eqnarray}
where 

\begin{eqnarray}\label{statefive}
\vert 0 \rangle_{L} & = & ( \vert 00000 \rangle + \vert 11111 \rangle )_{(0)} \otimes ( \vert 00000 \rangle + \vert 11111 \rangle )_{(1)} \otimes ( \vert 00000 \rangle + \vert 11111 \rangle )_{(2)} , \nonumber \\
\vert 1 \rangle_{L} & = & ( \vert 00000 \rangle - \vert 11111 \rangle )_{(0)}  \otimes ( \vert 00000 \rangle - \vert 11111 \rangle )_{(1)} \otimes ( \vert 00000 \rangle - \vert 11111 \rangle )_{(2)} , \nonumber \\
\vert 2 \rangle_{L} & = & ( \vert 00010 \rangle + \vert 11101 \rangle )_{(0)} \otimes ( \vert 00010 \rangle + \vert 11101 \rangle )_{(1)} \otimes ( \vert 00010 \rangle + \vert 11101 \rangle )_{(2)} , \nonumber \\
\vert 3 \rangle_{L} & = & ( \vert 00010 \rangle - \vert 11101 \rangle )_{(0)} \otimes ( \vert 00010 \rangle - \vert 11101 \rangle )_{(1)}  \otimes ( \vert 00010 \rangle - \vert 11101 \rangle )_{(2)} ,  \nonumber \\
\vdots & \nonumber \\
\vert 28 \rangle_{L} & = & ( \vert 11100 \rangle + \vert 00011 \rangle )_{(0)} \otimes ( \vert 11100 \rangle + \vert 00011 \rangle )_{(1)} \otimes ( \vert 11100 \rangle + \vert 00011 \rangle )_{(2)} ,  \nonumber \\
\vert 29 \rangle_{L} & = & ( \vert 11100 \rangle - \vert 00011 \rangle )_{(0)} \otimes ( \vert 11100 \rangle - \vert 00011 \rangle )_{(1)}  \otimes ( \vert 11100 \rangle - \vert 00011 \rangle )_{(2)} , \nonumber \\
\vert 30 \rangle_{L} & = & ( \vert 11110 \rangle + \vert 00001 \rangle )_{(0)} \otimes ( \vert 11110 \rangle + \vert 00001 \rangle )_{(1)} \otimes ( \vert 11110 \rangle + \vert 00001 \rangle )_{(2)} , \nonumber \\
\vert 31 \rangle_{L} & = & ( \vert 11110 \rangle - \vert 00001 \rangle )_{(0)} \otimes ( \vert 11110 \rangle - \vert 00001 \rangle )_{(1)} \otimes ( \vert 11110 \rangle - \vert 00001 \rangle )_{(2)}.
\end{eqnarray}

After applying the operator (\ref{openc5}), the three blocks that were into the computational basis have now been transformed into the GHZ basis.

To illustrate the interaction with the QEC and the workings of the restoring operation, we will now assume the following situation in which erasures may occur:

The encoded state $\vert \psi \rangle_{GHZ}$ in (\ref{fiveancil}) suffered erasure in the qubit $1$ of the index block $(0)$ and in the qubit 5 of the index block $(1)$, assume also the occurrence of phase-flip for these positions. After these erasures occur, the resulting state is as follows:


\begin{eqnarray}\label{fivedesc}
\vert \overline{\psi} \rangle_{GHZ} & = & \vert e_{0} \rangle  \otimes \vert \psi \rangle_{GHZ} \nonumber \\
 & = & \lambda_{0} \vert \overline{0} \rangle_{L} + \lambda_{1} \vert \overline{1} \rangle_{L} + \lambda_{2} \vert \overline{2} \rangle_{L} + \lambda_{3} \vert \overline{3} \rangle_{L} +  \cdots + \lambda_{28} \vert \overline{28} \rangle_{L}  \nonumber \\
          & & + \lambda_{29} \vert \overline{29} \rangle_{L} + \lambda_{30} \vert \overline{30} \rangle_{L} + \lambda_{31} \vert \overline{31} \rangle_{L},
\end{eqnarray}
where

\begin{eqnarray}\label{fiveerror}
\vert \overline{0} \rangle_{L} & = & ( \vert \overline{0}0000 \rangle - \vert \overline{1}1111 \rangle )_{(0)} \otimes ( \vert 0000\overline{0} \rangle - \vert 1111\overline{1} \rangle )_{(1)} \otimes ( \vert 00000 \rangle + \vert 11111 \rangle )_{(2)} , \nonumber \\
\vert \overline{1} \rangle_{L} & = & ( \vert \overline{0}0000 \rangle + \vert \overline{1}1111 \rangle )_{(0)} \otimes ( \vert 0000\overline{0} \rangle + \vert 1111\overline{1} \rangle )_{(1)}  \otimes ( \vert 00000 \rangle - \vert 11111 \rangle )_{(2)} , \nonumber \\
\vert \overline{2} \rangle_{L} & = & ( \vert \overline{0}0010 \rangle - \vert \overline{1}1101 \rangle )_{(0)} \otimes ( \vert 0001\overline{0} \rangle - \vert 1110\overline{1} \rangle )_{(1)} \otimes ( \vert 00010 \rangle + \vert 11101 \rangle )_{(2)} , \nonumber \\
\vert \overline{3} \rangle_{L} & = & ( \vert \overline{0}0010 \rangle + \vert \overline{1}1101 \rangle )_{(0)} \otimes ( \vert 0001\overline{0} \rangle + \vert 1110\overline{1} \rangle )_{(1)} \otimes ( \vert 00010 \rangle - \vert 11101 \rangle )_{(2)} ,  \nonumber \\
\vdots & & \nonumber \\
\vert \overline{28} \rangle_{L} & = & ( - \vert \overline{1}1100 \rangle + \vert \overline{0}0011 \rangle )_{(0)} \otimes ( \vert 1110\overline{0} \rangle - \vert 0001\overline{1} \rangle )_{(1)} \otimes ( \vert 11100 \rangle + \vert 00011 \rangle )_{(2)} , \nonumber \\
\vert \overline{29} \rangle_{L} & = & ( - \vert \overline{1}1100 \rangle - \vert \overline{0}0011 \rangle )_{(0)} \otimes ( \vert 1110\overline{0} \rangle - \vert 0001\overline{1} \rangle )_{(1)} \otimes ( \vert 11100 \rangle - \vert 00011 \rangle )_{(2)} , \nonumber \\
\vert \overline{30} \rangle_{L} & = & ( - \vert \overline{1}1110 \rangle + \vert \overline{0}0001 \rangle )_{(0)} \otimes ( \vert 1111\overline{0} \rangle - \vert 0000\overline{1} \rangle )_{(1)}  \otimes ( \vert 11110 \rangle + \vert 00001 \rangle )_{(2)} , \nonumber \\
\vert \overline{31} \rangle_{L} & = & ( - \vert \overline{1}1110 \rangle - \vert \overline{0}0001 \rangle )_{(0)} \otimes ( \vert 1111\overline{0} \rangle + \vert 0000\overline{1} \rangle )_{(1)} \otimes ( \vert 11110 \rangle - \vert 00001 \rangle )_{(2)} .
\end{eqnarray}

To extract the original message state $\ket{\psi}$, we must apply the restoring operation $\mathcal{R}$ (Theo\-rem \ref{restoper}), which for the present situation is given as follows:

\begin{eqnarray}\label{oprecov5}
\mathcal{R} & = & \bigg[ U_{rec}^{ 5, 1} \circ U_{dec} \Big( \vert \overline{\psi} \rangle_{GHZ} \otimes \ket{00000}_{(3)} \Big) \bigg] \bigg[ U_{rec}^{ 1, 0} \circ U_{dec} \Big( \vert \overline{\psi} \rangle_{GHZ} \otimes \ket{00000}_{(3)} \Big) \bigg]. 
\end{eqnarray} 

We first perform the $U_{dec}$ operator in  ($\vert \overline{\psi} \rangle_{GHZ} \otimes \ket{00000}_{(3)}$). Recall that this operator acts only in undamaged blocks. For each erroneous logical state in (\ref{fiveerror}), the only undamaged block has index $(2)$. In this case, the $U_{dec}$ operator is given as follows

\begin{eqnarray}\label{decod5}
U_{dec} & = & \prod_{d=0  (d \notin \{ 0, 1 \} )}^{2} \bigg( \prod_{i=1}^{5} C_{i(3),i(d)} \bigg) \prod_{d=0  (d \notin \{ 0, 1 \} )}^{2} \bigg( \prod_{i=1}^{5} C_{i(d),i(3)} H_{5(d)}  \prod_{i=1}^{4} C_{5(d),i(d)} \bigg) \nonumber \\
            & = & C_{1(3),1(2)} C_{2(3),2(2)} C_{3(3),3(2)} C_{4(3),4(2)} C_{5(3),5(2)} \nonumber \\
            &    & C_{1(2),1(3)} C_{2(2),2(3)} C_{3(2),3(3)} C_{4(2),4(3)} C_{5(2),5(3)}  \nonumber \\
            &    &  H_{5(2)}  C_{5(2),1(2)} C_{5(2),2(2)} C_{5(2),3(2)} C_{5(2),4(2)} .
\end{eqnarray}

Applying the decoding operator (\ref{decod5}) in  ($\vert \overline{\psi} \rangle_{GHZ} \otimes \ket{00000}_{(3)}$), we obtain

\begin{eqnarray}\label{fivedecerror}
\vert \overline{0} \rangle_{L} & = & ( \vert \overline{0}0000 \rangle - \vert \overline{1}1111 \rangle )_{(0)} \otimes ( \vert 0000\overline{0} \rangle - \vert 1111\overline{1} \rangle )_{(1)} \otimes  \vert 00000 \rangle_{(2)} \otimes  \vert 00000 \rangle_{(3)} , \nonumber \\
\vert \overline{1} \rangle_{L} & = & ( \vert \overline{0}0000 \rangle + \vert \overline{1}1111 \rangle )_{(0)} \otimes ( \vert 0000\overline{0} \rangle + \vert 1111\overline{1} \rangle )_{(1)}  \otimes  \vert 00000 \rangle_{(2)} \otimes  \vert 00001 \rangle_{(3)} , \nonumber \\
\vert \overline{2} \rangle_{L} & = & ( \vert \overline{0}0010 \rangle - \vert \overline{1}1101 \rangle )_{(0)} \otimes ( \vert 0001\overline{0} \rangle - \vert 1110\overline{1} \rangle )_{(1)} \otimes  \vert 00000 \rangle_{(2)} \otimes  \vert 00010 \rangle_{(3)} , \nonumber \\
\vert \overline{3} \rangle_{L} & = & ( \vert \overline{0}0010 \rangle + \vert \overline{1}1101 \rangle )_{(0)} \otimes ( \vert 0001\overline{0} \rangle + \vert 1110\overline{1} \rangle )_{(1)} \otimes  \vert 00000 \rangle_{(2)} \otimes  \vert 00011 \rangle_{(3)} , \nonumber \\
\vdots & & \nonumber \\
\vert \overline{28} \rangle_{L} & = & ( - \vert \overline{1}1100 \rangle + \vert \overline{0}0011 \rangle )_{(0)} \otimes ( \vert 1110\overline{0} \rangle - \vert 0001\overline{1} \rangle )_{(1)} \otimes  \vert 00000 \rangle_{(2)} \otimes  \vert 11100 \rangle_{(3)} , \nonumber \\
\vert \overline{29} \rangle_{L} & = & ( - \vert \overline{1}1100 \rangle - \vert \overline{0}0011 \rangle )_{(0)} \otimes ( \vert 1110\overline{0} \rangle - \vert 0001\overline{1} \rangle )_{(1)} \otimes  \vert 00000 \rangle_{(2)} \otimes  \vert 11101 \rangle_{(3)} , \nonumber \\
\vert \overline{30} \rangle_{L} & = & ( - \vert \overline{1}1110 \rangle + \vert \overline{0}0001 \rangle )_{(0)} \otimes ( \vert 1111\overline{0} \rangle - \vert 0000\overline{1} \rangle )_{(1)} \otimes  \vert 00000 \rangle_{(2)} \otimes  \vert 11110 \rangle_{(3)} , \nonumber \\
\vert \overline{31} \rangle_{L} & = & ( - \vert \overline{1}1110 \rangle - \vert \overline{0}0001 \rangle )_{(0)} \otimes ( \vert 1111\overline{0} \rangle + \vert 0000\overline{1} \rangle )_{(1)} \otimes  \vert 00000 \rangle_{(2)} \otimes  \vert 11111 \rangle_{(3)} .
\end{eqnarray}

Note that after applying $U_{dec}$, the block of index $(2)$: (a) was transformed from the GHZ basis into the computational basis; (b) was identically prepared in the block of index $(3)$; and, (c) had its $5$ qubits transformed into the $\ket{0}$ state.

We have that erasures occurred in the qubit of position $1$ of the index block $(0)$, and in the qubit of position $5$ of the index block $(1)$. So, for this situation, recovery operators (Theorem \ref{restoper}) are given as follows:

\begin{eqnarray}\label{recov510}
U_{rec}^{1, 0} & = &T_{1(3),5(3),4(0)} Z_{5(3), 4(0) } T_{1(3),5(3),4(0)} \prod_{i=1  (i \neq 1)}^{4} C_{i(3),i(0)} \prod_{i=1 (i \neq 1)}^{5} C_{1(3),i(0)}  \nonumber \\
& = & T_{1(3),5(3),4(0)} Z_{5(3), 4(0)} T_{1(3),5(3),4(0)} \nonumber \\
& & C_{2(3),2(0)}  C_{3(3),3(0)}  C_{4(3),4(0)}  \nonumber \\
&  & C_{1(3),2(0)} C_{1(3),3(0)}  C_{1(3),4(0)}  C_{1(3),5(0)}  
\end{eqnarray}  \\
where in this case, $\mathcal{W}_{(0)}= \{1, 2, 3, 4, 5 \} \setminus \{ 1 \} = \{ 2, 3, 4, 5 \}$, and $r = max_{r\neq 5}\{\mathcal{W}_{(0)} \}=4$; and also:

\begin{eqnarray}\label{recov551}
U_{rec}^{ 5, 1} & = & Z_{5(3), 4(1) }  \prod_{i=1}^{4} C_{i(3),i(1)} \nonumber \\ 
 & = & Z_{5(3), 4(1)} C_{1(3),1(1)} C_{2(3),2(1)} C_{3(3),3(1)}  C_{4(3),4(1)} 
\end{eqnarray}\\
where in this case, $\mathcal{W}_{(1)}= \{1, 2, 3, 4, 5 \} \setminus \{ 5 \} = \{ 1, 2, 3, 4 \}$, and $r = max_{r\neq 5}\{\mathcal{W}_{(1)}\}= 4$.  

Now, applying $U_{rec}^{1,0}$ in (\ref{fivedecerror}), we obtain

\begin{eqnarray}\label{recerror510}
\vert \overline{0} \rangle_{L} & = & ( \vert \overline{0}0000 \rangle - \vert \overline{1}1111 \rangle )_{(0)} \otimes ( \vert 0000\overline{0} \rangle - \vert 1111\overline{1} \rangle )_{(1)} \otimes  \vert 00000 \rangle_{(2)} \otimes  \vert 00000 \rangle_{(3)} , \nonumber \\
\vert \overline{1} \rangle_{L} & = & ( \vert \overline{0}0000 \rangle - \vert \overline{1}1111 \rangle )_{(0)} \otimes ( \vert 0000\overline{0} \rangle + \vert 1111\overline{1} \rangle )_{(1)}  \otimes  \vert 00000 \rangle_{(2)} \otimes  \vert 00001 \rangle_{(3)} , \nonumber \\
\vert \overline{2} \rangle_{L} & = & ( \vert \overline{0}0000 \rangle - \vert \overline{1}1111 \rangle )_{(0)} \otimes ( \vert 0001\overline{0} \rangle - \vert 1110\overline{1} \rangle )_{(1)} \otimes  \vert 00000 \rangle_{(2)} \otimes  \vert 00010 \rangle_{(3)} , \nonumber \\
\vert \overline{3} \rangle_{L} & = & ( \vert \overline{0}0000 \rangle - \vert \overline{1}1111 \rangle )_{(0)} \otimes ( \vert 0001\overline{0} \rangle + \vert 1110\overline{1} \rangle )_{(1)} \otimes  \vert 00000 \rangle_{(2)} \otimes  \vert 00011 \rangle_{(3)} ,  \nonumber \\
\vdots & & \nonumber \\
\vert \overline{28} \rangle_{L} & = & ( \vert \overline{0}0000 \rangle - \vert \overline{1}1111 \rangle )_{(0)} \otimes ( \vert 1110\overline{0} \rangle - \vert 0001\overline{1} \rangle )_{(1)} \otimes  \vert 00000 \rangle_{(2)} \otimes  \vert 11100 \rangle_{(3)} , \nonumber \\
\vert \overline{29} \rangle_{L} & = & ( \vert \overline{0}0000 \rangle - \vert \overline{1}1111 \rangle )_{(0)} \otimes ( \vert 1110\overline{0} \rangle - \vert 0001\overline{1} \rangle )_{(1)} \otimes  \vert 00000 \rangle_{(2)} \otimes  \vert 11101 \rangle_{(3)} , \nonumber \\
\vert \overline{30} \rangle_{L} & = & ( \vert \overline{0}0000 \rangle - \vert \overline{1}1111 \rangle )_{(0)} \otimes ( \vert 1111\overline{0} \rangle - \vert 0000\overline{1} \rangle )_{(1)} \otimes  \vert 00000 \rangle_{(2)} \otimes  \vert 11110 \rangle_{(3)} , \nonumber \\
\vert \overline{31} \rangle_{L} & = & ( \vert \overline{0}0000 \rangle - \vert \overline{1}1111 \rangle )_{(0)} \otimes ( \vert 1111\overline{0} \rangle + \vert 0000\overline{1} \rangle )_{(1)} \otimes  \vert 00000 \rangle_{(2)} \otimes  \vert 11111 \rangle_{(3)} .
\end{eqnarray}

Note that, in (\ref{recerror510}), the block of index $(0)$ is now in the form of a canonical GHZ state.

Applying $U_{rec}^{5,1}$ in (\ref{recerror510}), we obtain

\begin{eqnarray}\label{recerror551}
\vert \overline{0} \rangle_{L} & = & ( \vert \overline{0}0000 \rangle - \vert \overline{1}1111 \rangle )_{(0)} \otimes ( \vert 0000\overline{0} \rangle - \vert 1111\overline{1} \rangle )_{(1)} \otimes  \vert 00000 \rangle_{(2)} \otimes  \vert 00000 \rangle_{(3)} , \nonumber \\
\vert \overline{1} \rangle_{L} & = & ( \vert \overline{0}0000 \rangle - \vert \overline{1}1111 \rangle )_{(0)} \otimes ( \vert 0000\overline{0} \rangle - \vert 1111\overline{1} \rangle )_{(1)}  \otimes  \vert 00000 \rangle_{(2)} \otimes  \vert 00001 \rangle_{(3)} , \nonumber \\
\vert \overline{2} \rangle_{L} & = & ( \vert \overline{0}0000 \rangle - \vert \overline{1}1111 \rangle )_{(0)} \otimes ( \vert 0000\overline{0} \rangle - \vert 1111\overline{1} \rangle )_{(1)} \otimes  \vert 00000 \rangle_{(2)} \otimes  \vert 00010 \rangle_{(3)} , \nonumber \\
\vert \overline{3} \rangle_{L} & = & ( \vert \overline{0}0000 \rangle - \vert \overline{1}1111 \rangle )_{(0)} \otimes ( \vert 0000\overline{0} \rangle - \vert 1111\overline{1} \rangle )_{(1)} \otimes  \vert 00000 \rangle_{(2)} \otimes  \vert 00011 \rangle_{(3)} ,  \nonumber \\
\vdots & & \nonumber \\
\vert \overline{28} \rangle_{L} & = & ( \vert \overline{0}0000 \rangle - \vert \overline{1}1111 \rangle )_{(0)} \otimes ( \vert 0000\overline{0} \rangle - \vert 1111\overline{1} \rangle )_{(1)} \otimes  \vert 00000 \rangle_{(2)} \otimes  \vert 11100 \rangle_{(3)} , \nonumber \\
\vert \overline{29} \rangle_{L} & = & ( \vert \overline{0}0000 \rangle - \vert \overline{1}1111 \rangle )_{(0)} \otimes ( \vert 0000\overline{0} \rangle - \vert 1111\overline{1} \rangle )_{(1)} \otimes  \vert 00000 \rangle_{(2)} \otimes  \vert 11101 \rangle_{(3)} , \nonumber \\
\vert \overline{30} \rangle_{L} & = & ( \vert \overline{0}0000 \rangle - \vert \overline{1}1111 \rangle )_{(0)} \otimes ( \vert 0000\overline{0} \rangle - \vert 1111\overline{1} \rangle )_{(1)} \otimes  \vert 00000 \rangle_{(2)} \otimes  \vert 11110 \rangle_{(3)} , \nonumber \\
\vert \overline{31} \rangle_{L} & = & ( \vert \overline{0}0000 \rangle - \vert \overline{1}1111 \rangle )_{(0)} \otimes ( \vert 0000\overline{0} \rangle - \vert 1111\overline{1} \rangle )_{(1)} \otimes  \vert 00000 \rangle_{(2)} \otimes  \vert 11111 \rangle_{(3)} .
\end{eqnarray}

Observe that, in (\ref{recerror551}), the blocks of indices $(0)$ and $(1)$ are also now in the canonical GHZ state.

Therefore, after applying the operator (\ref{decod5}) and operators (\ref{recov510}) and (\ref{recov551}), the system and the environment will be in the state

\begin{eqnarray}\label{oprecov6}
& & \Big( \vert \overline{0}0000 \rangle - \vert \overline{1}1111 \rangle \Big)_{(0)} \otimes \Big( \vert 0000\overline{0} \rangle - \vert 1111\overline{1} \rangle \Big)_{(1)}  \otimes \Big( \ket{00000} \Big)_{(2)} \otimes \Big( \vert \psi \rangle \Big)_{(3)}.
\end{eqnarray}

Thus, the original message state $\ket{\psi}$ can be recovered free of erasures via the block of index $(3)$, despite $2$ erasures occurred when passing through the QEC.

It is important to note that to apply the recovery operators it is necessary to use the qubits of the index block $(3)$ (obtained from the undamaged block) and also the remaining qubits (not erased) of the index blocks $(0)$ and $(1)$. It means that, to recover the original state $\ket{\psi}$ via the block of index $(3)$, the collaboration of all blocks of the received state $\vert \overline{\psi} \rangle_{GHZ}$ is necessary . This concludes the example.

\section{Final Remarks}\label{conclus}

In this paper we presented a scheme for protecting $k$-qubit of information ($k \geq 3$) against $t = \lfloor k / 2 \rfloor$ erasures, since such that erasures occur on by distinct encoded blocks of k-qubit each.  This scheme improves the code proposed by Yang et al. \cite{Yang:schemGHZ3q} and makes use of $(t+1)$ redundant blocks in the GHZ basis. 

A special feature of the scheme presented is that no measurement is required, since information about the erasures is provided naturally by the system, for example, through spontaneous emission. This information can be captured by erasure detectors and lately treated via unitary operators that do not disturb the system. Another characteristic that should be remarked is that information can only be retrieved if there is a collaboration of all blocks that compose the state received.

The implementation of the proposed scheme is perfectly feasible, since it is achievable via unitary operators, which consist of an appropriate composition of quantum gates well-known in the literature (CNOT, Hadamard, Toffoli and $\sigma_{z}$-Pauli controled).

It is important to note that the operators that characterize the encoding operation (Theorem \ref{encapag}) and the restoring operation (Theorem \ref{restoper}) for this scheme can be adjusted to construct different quantum erasure-correcting codes. We must emphasize that the codes constructed via the proposed scheme can correct  only quantum erasures (i.e., changes which position is somehow flagged). Despite that, these codes can be concatenated with other codes such as quantum error-correcting codes to protect against the occurrence of computational errors \cite{Stace2009}.

Although the ratio $t / N$ decreases with $k$,\footnote{Here, $t =\lfloor k/2 \rfloor$ is the number of erasures, $N=k (\lfloor k/2 \rfloor + 1)$ is the total number of qubits required, and $k\geq 3$ is the codeword length.} \ we believe that the presented scheme can be useful in many applications as, for example, in the storage of quantum information for small-scale quantum computing, quantum information processing, and quantum communication. This is particularly emphasized  because the promising proposals of physical systems for quantum computers are based on the small-capacitance of  current technologies, such as: Josephson junctions \cite{Fazio1999, Zhou2005}; coupled quantum dots \cite{Loss1998,Caicedo2006}; neutral atoms in optical lattices \cite{Vala2005,Brennen1999}; and phosphorus dopants in silicon crystals \cite{Kane1998,Ladd2010}.

In future works, we suggest the application of the presented scheme in quantum information processing and quantum communication, such as in quantum secret sharing \cite{Hillery1999,Li2004} and in quantum cryptography \cite{Bennett1984,Gao2010}.


\section*{Acknowledgements}
We would like to thank A\'{e}rcio F. de Lima, Ello\'{a} B. Guedes and Givaldo O. dos Santos for  helpful discussions and also Markus Grassl for valuable comments. This work was partially supported from the Brazilian funding agencies CAPES/PIQDTEC and CNPq (309431/2006.9). 

\section*{References}

\addcontentsline{toc}{chapter}{Referências Bibliográficas}

%

\end{document}